\documentclass[a4paper,11pt]{article}
\usepackage{textcomp,amsmath,amsfonts,amssymb,amsthm,verbatim,mathrsfs}
\allowdisplaybreaks
\begin{document}

\title{On Completeness Results of Hoare Logic Relative to the Standard Model}
\author{
    Zhaowei Xu
    \footnotemark[1]
    \footnotemark[2],
    Wenhui Zhang
    \footnotemark[1],
    Yuefei Sui
    \footnotemark[3]
   }

\renewcommand{\thefootnote}{\fnsymbol{footnote}}

\footnotetext[1]{State Key Laboratory of Computer Science, Institute of Software, Chinese Academy of Sciences, Beijing, China}

\footnotetext[2]{Corresponding author at: University of Chinese Academy of Sciences, Beijing, China, xuzw@ios.ac.cn}

\footnotetext[3]{Key Laboratory of Intelligent Information Processing, Institute of Computing Technology, Chinese Academy of Sciences, Beijing, China}

\date{}

\maketitle

\begin{abstract}

The general completeness problem of Hoare logic relative to the standard model $N$ of Peano arithmetic has been studied by Cook, and it allows for the use of arbitrary arithmetical formulas as assertions. In practice, the assertions would be simple arithmetical formulas, e.g. of a low level in the arithmetical hierarchy. In addition, we find that, by restricting inputs to $N$, the complexity of the minimal assertion theory for the completeness of Hoare logic to hold can be reduced. This paper further studies the completeness of Hoare Logic relative to $N$ by restricting assertions to subclasses of arithmetical formulas (and by restricting inputs to $N$). Our completeness results refine Cook's result by reducing the complexity of the assertion theory.

\smallskip

\end{abstract}

\section{Introduction}

Hoare logic, first introduced by Hoare \cite{hoare_1} and further studied by Cook \cite{cook_1} and many other researchers, lays the foundation of program verification \cite{mirkowska_1,harel_1,reynolds_1}. For an introduction to Hoare logic, the reader should refer to \cite{apt_1,apt_2,kleymann_1,nipkow_1}. Let $L$ be the language of Peano arithmetic $PA$ \cite{model_1}, let $N$ be the standard model of $PA$, and let $Th(N)$ be the set of all true sentences in $N$. Hoare logic for the set $WP$ of while-programs with the language $L$ and the assertion theory $T\subset L$ is denoted $HL(T)$ \cite{bergstra_1}. The set $\{\varphi \in L : T \vdash \varphi \}$ of all theorems of $T\subset L$ is denoted $Thm(T)$. By Cook's completeness theorem, it follows that $Th(N)$ is the only extension $T$ of $PA$ such that $HL(T)$ is complete relative to $N$: for any $p,q\in L$ and $S\in WP$, if $N\models \{p\}S\{q\}$ then $HL(Th(N))\vdash \{p\}S\{q\}$; for any $T'\supseteq PA$ with $Thm(T') \subsetneqq Th(N)$ (note that $Thm(PA)\subsetneqq Th(N)$ follows from G\"{o}del's incompleteness theorem), there exist $p,q\in L$ and $S\in WP$ such that $N\models \{p\}S\{q\}$ but $HL(T')\not\vdash \{p\}S\{q\}$. Note that $Th(N)$ is not recursively enumerable, and even not arithmetical \cite[Lemma 17.3]{c. and l.}. That $Th(N)$ is the only extension of $PA$ for this completeness result to hold is due to the fact that it allows for the use of arbitrary arithmetical formulas as pre- and postconditions. In practice, the pre- and postconditions would be simple arithmetical formulas, e.g. of a low level in the arithmetical hierarchy \cite[Chapter IV]{soare_1}. The set $\{true, false\}$ of logical constants $true$ and $false$ is denoted $Cnt$. Apt's survey paper \cite[p 437]{apt_1} has shown that, for no recursively enumerable extension $T$ of $PA$, can $HL(T)$ derive all true Hoare's triples with pre- and postconditions restricted to $Cnt$. A natural question is whether there exists an arithmetical extension $T$ of $PA$ such that $HL(T)$ derives all true Hoare's triples with pre- and postconditions restricted to $Cnt$. Furthermore, we shall investigate the completeness of Hoare logic relative to $N$ with pre- and postconditions restricted to the arithmetical hierarchy.

Tennenbaum's theorem \cite{tennenbaum_1} says that addition and multiplication are not computable in nonstandard models. For practical purposes, it would be meaningless to consider computations over nonstandard models. By restricting inputs to the standard model (i.e. excluding nonstandard inputs), the discussion for computations over nonstandard models can be avoided. This idea has been applied to investigating the logical completeness of Hoare logic, and the complexity of the minimal assertion theory for the logical completeness to hold is reduced \cite{xu16}. Taking this idea into account, the technical lines of this paper are divided into two: one is including the nonstandard inputs; and the other is excluding the nonstandard inputs. For more details, the reader refers to Definition \ref{definition_of_completeness_of_Hoare_logic}. Therefore this paper deals with two completeness problems (completeness of Hoare logic relative to $N$ with pre- and postconditions restricted to $Cnt$ or the arithmetical hierarchy) in two technical lines (including or excluding nonstandard inputs).

This paper extends the conference version \cite{xu17}, dealing with completeness issues of Hoare logic relative to $N$ by restricting pre- and postconditions to $Cnt$ or the arithmetical hierarchy, without excluding nonstandard inputs.

The rest of this paper is organized as follows: the related work is given in Section 2; the basic preliminary results are presented in Section 3; completeness of $HL(T)$ with nonstandard inputs is shown in Section 4; completeness of $HL(T)$ without nonstandard inputs is shown in Section 5; comparison of $PA^*$, $PA^+$ and $PA \cup Tr^N(\Pi_1)$ is given in Section 6; Section 7 concludes the paper with a discussion of the results.

\section{Related work}

We call a set of assertions $A$ complete w.r.t. a class of programs $C$ if for any $p,q\in A$ and $S\in C$, whenever $\{p\} S \{q\}$ holds, then all intermediate assertions can be chosen from $A$. Apt et al. \cite{apt_3} studied the problem which sets of assertions are complete in the above sense. They have shown that $\Sigma_1$ is complete w.r.t. $WP$; $\Delta_1$ is not complete w.r.t. $WP$; and by allowing the use of an `auxiliary' coordinate, $\Delta_1$ is complete w.r.t. $WP$.

Clarke \cite{clarke_1} exhibited programming language structures for which Hoare logic is not complete relative to the finite structures, and observed that if a programming language possesses a relatively complete Hoare logic for partial correctness (relative to the finite structures) then the halting problem for finite interpretations must be decidable (Clarke's Observation). Lipton \cite{lipton_1}, Clarke et. al \cite{clarke_2}, and Grabowski \cite{grabowski_1} investigated under what circumstances the converse of Clarke's Observation holds. For the detailed relationship among their results, the reader refers to the Introduction of \cite{grabowski_1}. Note that their completeness results hold under the assumption that the halting problem for finite interpretations is decidable, whereas ours holds relative to the fixed structure $N$; and their axiom systems for Hoare logic are determined by the decision (or enumeration) procedures, while ours is given by Cook \cite{cook_1}.

Bergstra and Tucker \cite{bergstra_2} studied the logical completeness of Hoare logic with nonstandard inputs: $Th(N)$ is the only extension $T$ of $PA$ such that $HL(T)$ is logically complete. Xu et al. \cite{xu16} studied the logical completeness of Hoare logic without nonstandard inputs: $PA^+$ (cf. Definition \ref{definition_of_pa_plus}) is the minimal extension $T$ of $PA$ such that $HL(T)$ is logically complete when inputs range over $N$. To establish this completeness result, the technical line of reducing from $HL(T)$ to $T$ (cf. Subsection \ref{reduction_from_hl_to_pa}) has been adopted, which will also be followed in this paper. Kozen and Tiuryn \cite{kozen_1} investigated the completeness of propositional Hoare logic with assertions and programs abstracted to propositional symbols.

\section{Preliminaries}

First some notations are introduced: in syntax, we write $\neg$, $\wedge$, $\vee$, $\rightarrow$, $\leftrightarrow$, $\forall$, $\exists$ to denote the negation, conjunction, disjunction, conditional, biconditional connectives and the universal, existential quantifiers; in semantics, we write $\sim$, $\&$, $|$, $\Rightarrow$, $\Leftrightarrow$, $\textbf{A}$, $\textbf{E}$ to denote the corresponding connectives and quantifiers.

\subsection{Peano arithmetic}

Let $\Sigma = \{ 0,1,+,\cdot,< \}$ be the signature of $L$. For simplicity, the sum of $1$ with itself $n$ times is abbreviated $n$. We use $n$ to denote both a closed term and a natural number, and use $M$ to denote both a model and its domain, which will be clear from the context. The formula $\varphi(t/x)$ stands for the result of simultaneously substituting $t$ for the free occurrences of $x$ in $\varphi$; and $\varphi(t/x)$ will be denoted $\varphi(t)$ if the default variable $x$ is obvious. The denotation of a term $t$ at an assignment $v$ (for all the first order variables) over a model $M$ of $L$, denoted $t^{M,v}$, receives the standard meaning. The satisfaction of a formula $\varphi\in L$ at an assignment $v$ over a model $M$ of $L$, denoted $M,v \models \varphi$, is defined as usual; the satisfaction of $\varphi$ in $M$, denoted $M\models \varphi$, is defined such that for any assignment $v$ over $M$, $M,v\models \varphi$; the satisfaction of a theory $T\subset L$ in $M$, denoted $M\models T$, is defined such that for any $\varphi\in T$, $M\models \varphi$; the satisfaction of $\varphi$ in a theory $T\subset L$, denoted $T\models \varphi$, is defined such that for any $M\models T$, $M\models \varphi$. And the derivation of a formula $\varphi\in L$ from a theory $T\subset L$, denoted $T\vdash \varphi$, is defined as usual. Besides the standard model $N$, $PA$ has nonstandard models $M$: $M$ has a standard part $N^M$ which is isomorphic to $N$; each element of $N^M$ is denoted $n$ as well. The distinguished axiom of $PA$ is the induction axiom scheme $\varphi(0,\vec{y})\wedge \forall x\big(\varphi(x,\vec{y})\rightarrow \varphi(x+1,\vec{y})\big)\rightarrow \forall x\ \varphi(x,\vec{y})$, where $\varphi(x,\vec{y}) \in L$. From $PA$, one can deduce the least-number principle $\exists x\ \varphi(x,\vec{y})\rightarrow \exists z ( \varphi(z,\vec{y})\wedge \forall u<z\ \neg \varphi(u,\vec{y}) )$, where $\varphi(x,\vec{y}) \in L$.

Generalized $\Sigma_n$-formulas and generalized $\Pi_n$-formulas of $L$ are defined as follows: a generalized $\Sigma_0$-formula (or a generalized $\Pi_0$-formula) is a formula built up from atomic formulas using only negation, conjunction, disjunction, and bounded quantifications $\forall x<t$ and $\exists x<t$, where $t$ is a term of $L$; a generalized $\Sigma_{n+1}$-formula is a formula obtainable from generalized $\Pi_n$-formulas by conjunction, disjunction, bounded quantifications, and unbounded existential quantification; a generalized $\Pi_{n+1}$-formula is a formula obtainable from generalized $\Sigma_n$-formulas by conjunction, disjunction, bounded quantifications and unbounded universal quantification. $\Sigma_n$-formulas and $\Pi_n$-formulas of $L$ are defined as follows: a $\Sigma_0$-formula (or a $\Pi_0$-formula) is a generalized $\Sigma_0$-formula; a $\Sigma_{n+1}$-formula is a formula of the form $\exists x\ \psi$ with $\psi$ being a $\Pi_n$-formula; a $\Pi_{n+1}$-formula is a formula of the form $\forall x\ \psi$ with $\psi$ being a $\Sigma_n$-formula. The set of all $\Sigma_n$-formulas is denoted $\Sigma_n$, and similarly for $\Pi_n$. (Generalized) $\Sigma_n$-sentences are (generalized) $\Sigma_n$-formulas without free variables, and similarly for (generalized) $\Pi_n$-sentences. The set of all true $\Sigma_n$-sentences in $N$ is denoted $Tr^N(\Sigma_n)$, and similarly for $Tr^N(\Pi_n)$.

It holds, in $PA$, that every generalized $\Sigma_n$-formula (resp. generalized $\Pi_n$-formula) is logically equivalent to a $\Sigma_n$-formula (resp. $\Pi_n$-formula). For the membership relation $\in$, besides the standard meaning, we sometimes adopt a nonstandard meaning: by $\varphi\in A$ (the nonstandard meaning) is meant that there exists $\psi\in A$ (the standard meaning) such that $PA\vdash\varphi\leftrightarrow \psi$. Only when the standard meaning of $\in$ is inapplicable, can the nonstandard meaning be adopted. The reader should keep this in mind. Then $\varphi\in\Sigma_n$ implies $\neg\varphi\in\Pi_n$, and $\varphi\in\Pi_n$ implies $\neg\varphi\in\Sigma_n$. Both $\Sigma_n$ and $\Pi_n$ are closed under conjunction and disjunction. For any $i \geq 0$, $\Sigma_i, \Pi_i\subset \Sigma_{i+1},\Pi_{i+1}$, and $\Sigma_i\nsubseteq \Pi_i$, $\Pi_i\nsubseteq \Sigma_i$. For the truth of these results, the reader refers to \cite[Chapter IV]{soare_1}.

We say that a set of natural numbers is $\Sigma_n$ (resp. $\Pi_n$) if it is arithmetically definable (or arithmetical for short) by a $\Sigma_n$-formula (resp. by a $\Pi_n$-formula); a set of natural numbers is $\Delta_n$ if it is both $\Sigma_n$ and $\Pi_n$. Note that a set of natural numbers is recursively enumerable (or r.e. for short) iff it is $\Sigma_1$, and that a set of natural numbers is recursive iff it is $\Delta_1$ \cite[Section 7.2]{c. and l.}. Theorem 16.13 in \cite{c. and l.} says that for all $\Sigma_1$-sentences $\varphi$, $N\models \varphi$ iff $PA\vdash \varphi$. Let $\ulcorner\varphi\urcorner$ be a fixed G\"{o}del's numbering function \cite[Chapter 15]{c. and l.}. By arithmetical definability of the theory $T\subset L$ is meant that the set $\{\ulcorner\varphi\urcorner : \varphi\in T\}$ of natural numbers is arithmetical. G\"{o}del's diagonal lemma \cite[Lemma 17.1]{c. and l.} says that for any $T \supseteq PA$ and any $\varphi(x)\in L$ there is a sentence $G\in L$ such that $T\vdash G\leftrightarrow \varphi(\ulcorner G\urcorner)$.

\subsection{Hoare logic}

Based on the language $L$, together with the program constructs $\{$ $:=$, $;$, $if$, $then$, $else$, $fi$, $while$, $do$, $od$ $\}$, a while-program $S$ is defined by $S ::= x := E \mid S_1;S_2 \mid if\ B\ then\ S_1\ else\ S_2\ fi \mid while\ B\ do\ S_0\ od$, where an expression $E$ is defined by $E ::= 0 \mid 1 \mid x \mid E_1+E_2 \mid E_1\cdot E_2$, and a boolean expression $B$ is defined by $B ::= E_1 < E_2 \mid \neg B_1 \mid B_1\rightarrow B_2$. The set of all such while-programs is denoted $WP$. The set of all assignment programs $x := E$ is denoted $AP$. For $S\in WP$, the vector $(x_1, x_2, \ldots, x_m)$ of all $m$ program variables $x_1,$ $x_2,$ $\ldots,$ $x_m$ occurring in $S$ will be denoted $\vec{x}$; the vector $(n_1,n_2,\ldots,n_m)$ of $m$ natural numbers $n_1,$ $n_2,$ $\ldots,$ $n_m\in N$ will be denoted $\vec{n}$; the connectives will be assumed to distribute over the components of the vectors (for instance, $\vec{n}\in N$ means $n_1,$ $n_2,$ $\ldots,$ $n_m\in N$, and $\vec{x} = \vec{n}$ means $\bigwedge_{i = 1}^{m} x_i = n_i$). Let the program variables considered below occur among $\vec{x}$, the vector of all program variables of the target program. For a model $M$ of $L$, let $v$ be an assignment over $M$ for all the first order variables (including $\vec{x}$), let $v(\vec{x})$ be the vector of elements of $M$ assigned to $\vec{x}$ at $v$, and let $v(\vec{a}/\vec{x})$ be an assignment as $v$ except that $v(\vec{a}/\vec{x})(\vec{x}) = \vec{a}$.

For every $S\in WP$ and every model $M$ of $L$, the input-output relation $R_S^M$ of $S$ in $M$ is a binary relation on the set of all assignments over $M$ inductively defined as follows:
\begin{itemize}
  \item $(v,v')\in R_{x:=E}^{M}$ $\Leftrightarrow$ $v'=v(E^{M,v}/x)$, where $E^{M,v}$ receives the standard meaning;
  \item $(v,v')\in R_{S_1;S_2}^{M}$ $\Leftrightarrow$ $(v,v')\in R_{S_1}^{M}\circ R_{S_2}^{M}$, where $(z,z')\in R_1\circ R_2$ $\Leftrightarrow$ $\textbf{E} z'' ( (z,z'')\in R_1$ $\&$ $(z'',z')\in R_2 )$;
  \item $(v,v')\in R_{if\ B\ then\ S_1\ else\ S_2\ fi}^{M}$ $\Leftrightarrow$ $M,v\models B$ $\&$ $(v,v')\in R_{S_1}^{M}$ $|$ $M,v\not\models B$ $\&$ $(v,v')\in R_{S_2}^{M}$;
  \item $(v,v')\in R_{while\ B\ do\ S_{0}\ od}^{M}$ $\Leftrightarrow$ $\textbf{E} i\in N$, $\textbf{E}\vec{a_0},\ldots,\vec{a_i} \in M$ $( v(\vec{x}) = \vec{a_0}$ $\&$ $\textbf{A} j < i ( M,v(\vec{a_j}/\vec{x})\models B$ $\&$ $(v(\vec{a_j}/\vec{x}), v(\vec{a_{j+1}}/\vec{x}))\in R_{S_0}^{M} ) $ $\&$ $v' = v(\vec{a_i}/\vec{x})$ $\&$ $M,v'\not\models B )$.
\end{itemize}

Given $S\in WP$ and a model $M$ of $L$, $R_S^M$ defines in $M$ a vectorial function $\vec{y} = f_S^M(\vec{x})$ such that for every $\vec{a},\vec{b}\in M$, $f_S^{M}(\vec{a}) = \vec{b}$ iff $\textbf{E} v,v' ( v(\vec{x})=\vec{a}\ \&\ v'(\vec{x})=\vec{b}\ \&\ (v,v')\in R_S^M )$. Given a model $M$ of $L$ and an asserted program $\{p\}S\{q\}$, $\{p\}S\{q\}$ is satisfied at $M$, denoted $M\models \{p\}S\{q\}$, iff $\textbf{A} v [ M,v\models p \Rightarrow \textbf{A}v'( (v,v')\in R_S^M \Rightarrow M,v'\models q ) ]$. Given a theory $T\subset L$ and an asserted program $\{p\}S\{q\}$, $\{p\}S\{q\}$ is satisfied at $T$, denoted $HL(T)\models \{p\}S\{q\}$, iff $\textbf{A} M (M\models T \Rightarrow M \models \{p\}S\{q\})$. $HL(T)$ has the usual axiom system \cite{bergstra_1}; the derivability of $\{p\}S\{q\}$ in $HL(T)$ is denoted $HL(T)\vdash \{p\}S\{q\}$. By the logical completeness of $HL(T)$ we mean that for all asserted programs $\{p\}S\{q\}$, $HL(T)\vdash \{p\}S\{q\}$ iff $HL(T)\models \{p\}S\{q\}$.

\theoremstyle{definition}
\newtheorem{definition_of_logical_completeness_of_hl_without_nonstandard_inputs}{Definition}[subsection]
\begin{definition_of_logical_completeness_of_hl_without_nonstandard_inputs}[{cf. \cite[Definition 1.1]{xu16}}]\label{definition_of_logical_completeness_of_hl_without_nonstandard_inputs}
   $HL(T)$ is logically complete when inputs range over $N$ if for every $S\in WP$ with program variables $\vec{x}$, every $p,q\in L$ ($p,$ $q$ could contain other first-order variables than those in $\vec{x}$), and every $\vec{n}\in N$, $HL(T)\vdash \{p \wedge \vec{x} = \vec{n}\} S \{q\}$ iff $HL(T)\models \{p \wedge \vec{x} = \vec{n}\} S \{q\}$.
\end{definition_of_logical_completeness_of_hl_without_nonstandard_inputs}

\newtheorem{definition_of_completeness_of_Hoare_logic}[definition_of_logical_completeness_of_hl_without_nonstandard_inputs]{Definition}
\begin{definition_of_completeness_of_Hoare_logic}\label{definition_of_completeness_of_Hoare_logic}
Let $P$ and $Q$ denote respectively the levels of choices of preconditions and postconditions (i.e. $Cnt$ or $\Sigma_i$, $\Pi_i$, $i\geq 0$), and let $R$ denote the sets of programs (i.e. $AP$ or $WP$).

(i) $HL(T)$ is complete relative to $N$ for $\{P\}R\{Q\}$ (with nonstandard inputs) if for any $p\in P$, $S\in R$, and $q\in Q$, $N\models \{p\} S \{q\}$ implies $HL(T)\vdash \{p\} S \{q\}$;

(ii) $HL(T)$ is complete relative to $N$ for $\{P\}R\{Q\}$ without nonstandard inputs if for any $S\in R$ with program variables $\vec{x}$, $p(\vec{u},\vec{x})\in P$, $q(\vec{u},\vec{x})\in Q$ (besides $\vec{x}$, $p$ and $q$ could contain other free variables $\vec{u}$), and $\vec{m},\vec{n}\in N$, $N\models \{p \wedge (\vec{u},\vec{x}) = (\vec{m},\vec{n})\} S \{q\}$ implies $HL(T)\vdash \{p \wedge (\vec{u},\vec{x}) = (\vec{m},\vec{n})\} S \{q\}$.
\end{definition_of_completeness_of_Hoare_logic}

Note that in Definition \ref{definition_of_completeness_of_Hoare_logic} (ii), we restrict both the inputs of program variables and the inputs of other free variables to $N$; while in Definition \ref{definition_of_logical_completeness_of_hl_without_nonstandard_inputs}, we only restrict the inputs of program variables to $N$.

\subsection{Reduction from $HL(T)$ to $T$}\label{reduction_from_hl_to_pa}

Let $\langle x, y\rangle$, $L(z)$ and $R(z)$ be the pairing functions with $\langle L(z),R(z)\rangle=z$, $L(\langle x,y\rangle)=x$ and $R(\langle x,y\rangle)=y$ \cite[Theorem 2.1]{computability_1}. For notational convenience, we denote $(L(z),R(z))$ by $\overline{z}$. The functions $\langle x, y\rangle$ and $\overline{z}$ can be extended to $n$-tuples (for each $n\in N$) by setting $\langle x_1,x_2,\ldots,x_n\rangle = \langle x_1,\langle x_2,\ldots,x_n\rangle\rangle$ and $\overline{\langle x_1,x_2,\ldots,x_n\rangle} = (x_1,\overline{\langle x_2,\ldots,x_n\rangle})$. Let $(x)_i$ be G\"{o}del's $\beta$-function such that for each finite sequence $a_0,a_1,\ldots,a_n$ of natural numbers, there exists a natural number $w$ such that $(w)_i = a_i$ for all $i\leq n$ \cite[Theorem 2.4]{computability_1}. Note that the graph relations of these functions are all $\Sigma_1$.

\theoremstyle{plain}
\newtheorem{pairing_functions}{Lemma}[section]
\begin{pairing_functions}[{cf. \cite[p45]{computability_1}}]\label{pairing_functions}
$PA$ proves that

(a) $\langle L(z),R(z)\rangle=z$;

(b) $L(\langle x,y\rangle)=x$;

(c) $R(\langle x,y\rangle)=y$.
\end{pairing_functions}

\newtheorem{beta_function}[pairing_functions]{Lemma}
\begin{beta_function}[{cf. \cite[p63]{model_1}}]\label{beta_function}
$PA$ proves that

(a) $\forall x \exists y\ (y)_0=x$;

(b) $\forall x,y,z \exists w (\forall i<z\ (w)_i=(y)_i \wedge (w)_z=x)$.
\end{beta_function}

\theoremstyle{definition}
\newtheorem{definition_of_alpha_S}{Definition}[subsection]
\begin{definition_of_alpha_S}[{The definition of $\alpha_S$, cf. \cite[Definition 3.1.1]{xu16}}]\label{definition_of_alpha_S}
For every $S\in WP$ with program variables $\vec{x}$, the generalized $\Sigma_1$-formula $\alpha_S(\vec{x},\vec{y}) \in L$, where $\vec{y} = (y_1, y_2, \ldots, y_m)$ is disjoint from $\vec{x} = (x_1, x_2, \ldots, x_m)$, is defined inductively as follows.

Assignment: $S \equiv x_i := E$
\begin{equation*}
\alpha_S(\vec{x},\vec{y}) ::= y_i=E(\vec{x}) \wedge \bigwedge_{1 \leq j \leq m}^{j\neq i} y_j=x_j;
\end{equation*}

Composition: $S \equiv S_1;S_2$
\begin{equation*}
\alpha_S(\vec{x},\vec{y}) ::= \exists \vec{z} (\alpha_{S_1}(\vec{x},\vec{z}/\vec{y})\wedge \alpha_{S_2}(\vec{z}/\vec{x},\vec{y}));
\end{equation*}

Conditional: $S \equiv if\ B\ then\ S_1\ else\ S_2\ fi$
\begin{equation*}
\alpha_S(\vec{x},\vec{y}) ::= (B(\vec{x})\wedge \alpha_{S_1}(\vec{x},\vec{y})) \vee (\neg B(\vec{x}) \wedge \alpha_{S_2}(\vec{x},\vec{y}));
\end{equation*}

Iteration: $S \equiv while\ B\ do\ S_0\ od$. We first let
\begin{eqnarray*}
  A_S(i,w,\vec{x},\vec{y}) &::=& \vec{x}=\overline{(w)_0} \wedge \forall j<i (B(\overline{(w)_j}/\vec{x}) \\
                         &&       \wedge\: \alpha_{S_0}(\overline{(w)_j}/\vec{x},\overline{(w)_{j+1}}/\vec{y})) \wedge \vec{y} = \overline{(w)_i}
\end{eqnarray*}
then set
\begin{equation*}
\alpha_S^*(i,\vec{x},\vec{y}) ::= \exists w\ A_S(i,w,\vec{x},\vec{y})
\end{equation*}
and finally define
\begin{equation*}
\alpha_S(\vec{x},\vec{y}) ::= \exists i\ \alpha_S^*(i,\vec{x},\vec{y}) \wedge \neg B(\vec{y}/\vec{x}).
\end{equation*}
\end{definition_of_alpha_S}

\newtheorem{alpha_defines_S}[definition_of_alpha_S]{Lemma}
\begin{alpha_defines_S}[{Arithmetical definability of recursive functions, cf. \cite[Lemma 3.1.2]{xu16}}] \label{alpha_defines_S}
  For every $S\in WP$ and every $\vec{a},\vec{b}\in N$, $f_S^N(\vec{a}) = \vec{b}$ iff $N\models \alpha_S(\vec{a},\vec{b})$.
\end{alpha_defines_S}

\theoremstyle{plain}
\newtheorem{from_hl_to_pa}[definition_of_alpha_S]{Theorem}
\begin{from_hl_to_pa}[{Reduction from $HL(T)$ to $T$, cf. \cite[Theorem 3.1.3]{xu16}}]\label{from_hl_to_pa}
For every $PA\subseteq T \subseteq Th(N)$, every $p,q\in L$ and every $S\in WP$,
\begin{equation*}
  HL(T)\vdash \{p\} S \{q\} \mbox{\ iff\ } T \vdash p(\vec{x})\wedge \alpha_S(\vec{x},\vec{y})\rightarrow q(\vec{y}/\vec{x}).
\end{equation*}
\end{from_hl_to_pa}

\newtheorem{completeness_of_Hoare_logic_for_Cnt_AP_Cnt}[definition_of_alpha_S]{Corollary}
\begin{completeness_of_Hoare_logic_for_Cnt_AP_Cnt}\label{completeness_of_Hoare_logic_for_Cnt_AP_Cnt}
$HL(PA)$ is complete relative to $N$ for $\{Cnt\}AP\{Cnt\}$.
\end{completeness_of_Hoare_logic_for_Cnt_AP_Cnt}
\begin{proof}
  Immediate from Definition \ref{definition_of_completeness_of_Hoare_logic} (i) and Theorem \ref{from_hl_to_pa}.
\end{proof}

\section{Completeness of $HL(T)$ for $\{P\}WP\{Q\}$ relative to $N$ (with nonstandard inputs)}

This section devotes to studying the completeness of $HL(T)$ for $\{P\}WP\{Q\}$ relative to $N$ (with nonstandard inputs). In Subsection \ref{when_P_and_Q_are_equal_to_Cnt_with_nonstandard_inputs}, the case when $P,Q$ $=$ $Cnt$ is investigated. To investigate the case when $P,Q$ $=$ $\Sigma_i, \Pi_i$, $i\geq 0$, we remark that if $P$ or $Q$ is expanded to a larger level in the arithmetical hierarchy, then $T$ will correspondingly be expanded to ``a larger level in the hierarchy of $Th(N)$''. Hence the hierarchy of $Th(N)$ will be studied: whether $Tr^N(\Sigma_{n+1})$ and $Tr^N(\Pi_{n+1})$ can be derived from $PA \cup Tr^N(\Pi_n)$. In Subsection \ref{when_P_and_Q_are_equal_to_Cnt_with_nonstandard_inputs}, the case when $P,Q$ $=$ $Cnt$, is investigated. In Subsection \ref{hierarchy of_Th_N}, the hierarchy of $Th(N)$ is given. In Subsection \ref{when_P_and_Q_range_over_the_arithmetical_hierarchy_with_nonstandard_inputs}, the case when $P,Q$ $=$ $\Sigma_i, \Pi_i$, $i\geq 0$, is investigated.

\subsection{When $P, Q$ $=$ $Cnt$}\label{when_P_and_Q_are_equal_to_Cnt_with_nonstandard_inputs}

\theoremstyle{plain}
\newtheorem{unprovability_of_the_halting_problem_for_all_inputs}{Lemma}[subsection]
\begin{unprovability_of_the_halting_problem_for_all_inputs}\label{unprovability_of_the_halting_problem_for_all_inputs}
  There exists $S\in WP$ such that $N\models \forall \vec{x},\vec{y} \neg \alpha_S(\vec{x},\vec{y})$ and $PA\nvdash \forall \vec{x},\vec{y} \neg \alpha_S(\vec{x},\vec{y})$.
\end{unprovability_of_the_halting_problem_for_all_inputs}
\begin{proof}
Note that the set of Hoare's triples $\{\{true\} S \{false\} : S\in WP, N\models \{true\} S \{false\} \}$ represents the complement of the halting problem, and hence is not r.e. (cf. the Fact in \cite[p 437]{apt_1}). On the other hand, the set of Hoare's triples $\{\{true\} S \{false\} : S\in WP, HL(PA)\vdash \{true\} S \{false\} \}$ is r.e. By soundness of Hoare logic, it follows that $\{\{true\} S \{false\} : S\in WP, HL(PA)\vdash \{true\} S \{false\} \}$ $\subsetneqq$ $\{\{true\} S \{false\} : S\in WP, N\models \{true\} S \{false\} \}$. Then there exists $S\in WP$ such that $N\models \{true\} S \{false\}$ but $HL(PA)\nvdash \{true\} S \{false\}$. By Lemma \ref{alpha_defines_S}, jointly with Theorem \ref{from_hl_to_pa}, it follows that there exists $S\in WP$ such that $N\models \forall \vec{x},\vec{y} \neg \alpha_S(\vec{x},\vec{y})$ and $PA\nvdash \forall \vec{x},\vec{y} \neg \alpha_S(\vec{x},\vec{y})$.
\end{proof}

\theoremstyle{definition}
\newtheorem{definition_of_pa_star}[unprovability_of_the_halting_problem_for_all_inputs]{Definition}
\begin{definition_of_pa_star}[The definition of $PA^*$]\label{definition_of_pa_star}
  We define $PA^*$ to be
  \begin{eqnarray*}
    PA^* &::=& PA \cup \{ \forall \vec{x},\vec{y} \neg \alpha_S(\vec{x},\vec{y}) : S\in WP \\
     && \&\ N\models \forall \vec{x},\vec{y} \neg \alpha_S(\vec{x},\vec{y})\ \&\ PA\nvdash \forall \vec{x},\vec{y} \neg \alpha_S(\vec{x},\vec{y})\}.
  \end{eqnarray*}
\end{definition_of_pa_star}

\theoremstyle{plain}
\newtheorem{property_of_pa_star}[unprovability_of_the_halting_problem_for_all_inputs]{Lemma}
\begin{property_of_pa_star}\label{property_of_pa_star}
It is the case that

(i) $Thm(PA)$ $\subsetneqq$ $Thm(PA^*)$ $\subseteq$ $Thm(PA \cup Tr^N(\Pi_1))$;

(ii) $PA^*$ and $Thm(PA^*)$ are $\Sigma_2$.
\end{property_of_pa_star}
\begin{proof}
(i) $Thm(PA)$ $\subsetneqq$ $Thm(PA^*)$ follows from Lemma \ref{unprovability_of_the_halting_problem_for_all_inputs} and Definition \ref{definition_of_pa_star}. $Thm(PA^*)$ $\subseteq$ $Thm(PA \cup Tr^N(\Pi_1))$ follows from Definition \ref{definition_of_pa_star}, together with the fact that $\forall \vec{x},\vec{y} \neg \alpha_S(\vec{x},\vec{y})$ is logically equivalent to a $\Pi_1$-sentence.

(ii) Since $Thm(PA^*)$ is r.e. in $PA^*$, i.e. $\Sigma_1$ in $PA^*$, to prove $Thm(PA^*)$ is $\Sigma_2$, it suffices to prove that $PA^*$ is $\Sigma_2$. Consider the statement $\varphi\in PA^*$ as follows: by definition of $PA^*$, it is equivalent to saying that $\varphi\in PA$, or there exists $S\in WP$ such that $\varphi = \forall \vec{x},\vec{y} \neg \alpha_S(\vec{x},\vec{y})$, $N\nvDash \neg\varphi$ and $PA\nvdash \varphi$; since $\neg\forall \vec{x},\vec{y} \neg \alpha_S(\vec{x},\vec{y})$ is logically equivalent to a $\Sigma_1$-sentence, and a $\Sigma_1$-sentence is true in $N$ iff it is a theorem of $PA$, it is equivalent to saying that $\varphi\in PA$, or there exists $S\in WP$ such that $\varphi = \forall \vec{x},\vec{y} \neg \alpha_S(\vec{x},\vec{y})$, $\neg\varphi \not\in Thm(PA)$ and $\varphi\not\in Thm(PA)$. Note that the set $\{\varphi: \varphi = \forall \vec{x},\vec{y} \neg \alpha_S(\vec{x},\vec{y})\ \&\ S\in WP\}$ is $\Delta_1$ and hence $\Sigma_2$. Since $Thm(PA)$ is $\Sigma_1$, we have that the set $\{\varphi: \varphi \not\in Thm(PA)\}$ is $\Pi_1$ and hence $\Sigma_2$, and the set $\{\varphi: \neg\varphi \not\in Thm(PA)\}$ is $\Pi_1$ and hence $\Sigma_2$. By closure of $\Sigma_2$ under conjunction, it follows that the set $\{ \varphi: \varphi = \forall \vec{x},\vec{y} \neg \alpha_S(\vec{x},\vec{y})\ \&\ S\in WP\ \&\ \neg\varphi \not\in Thm(PA)\ \&\ \varphi\not\in Thm(PA)\}$ is $\Sigma_2$. Moreover, since $PA$ is $\Delta_1$, we have that the set $\{\varphi: \varphi\in PA\}$ is $\Sigma_2$. By closure of $\Sigma_2$ under disjunction, it follows that $PA^*$ is $\Sigma_2$.
\end{proof}

\theoremstyle{definition}
\newtheorem{minimal_extension_for_the_completeness_of_Hoare_logic}[unprovability_of_the_halting_problem_for_all_inputs]{Definition}
\begin{minimal_extension_for_the_completeness_of_Hoare_logic}\label{minimal_extension_for_the_completeness_of_Hoare_logic}
  $T'$ is the minimal extension $T$ of $PA$ such that the property $p(T)$ of $T$ holds if

  (i) $p(T')$ holds; and

  (ii) for any $T''\supseteq PA$ with $Thm(T'')\subsetneqq Thm(T')$, $p(T'')$ doesn't hold.
\end{minimal_extension_for_the_completeness_of_Hoare_logic}

\theoremstyle{plain}
\newtheorem{completeness_of_Hoare_logic_for_Cnt_WP_Cnt_with_nonstandard_inputs}[unprovability_of_the_halting_problem_for_all_inputs]{Theorem}
\begin{completeness_of_Hoare_logic_for_Cnt_WP_Cnt_with_nonstandard_inputs}\label{completeness_of_Hoare_logic_for_Cnt_WP_Cnt_with_nonstandard_inputs}
  $PA^*$ is the minimal extension $T$ of $PA$ such that $HL(T)$ is complete relative to $N$ for $\{Cnt\}WP\{Cnt\}$ with nonstandard inputs.
\end{completeness_of_Hoare_logic_for_Cnt_WP_Cnt_with_nonstandard_inputs}
\begin{proof}
We first show that $HL(PA^*)$ is complete relative to $N$ for $\{Cnt\}WP\{Cnt\}$ with nonstandard inputs. By Definition \ref{definition_of_completeness_of_Hoare_logic} (i), we have to prove that for any $p,q\in Cnt$, and $S\in WP$, $N\models \{p\} S \{q\}$ implies $HL(PA^*)\vdash \{p\} S \{q\}$. Let $N\models \{p\} S \{q\}$ with $p,q\in Cnt$ and $S\in WP$. It remains to prove that $HL(PA^*)\vdash \{p\} S \{q\}$. For $p\equiv false$ or $q \equiv true$, it's easy to see that $PA^* \vdash p(\vec{x})\wedge \alpha_S(\vec{x},\vec{y})\rightarrow q(\vec{y}/\vec{x})$; by Theorem \ref{from_hl_to_pa}, it follows that $HL(PA \cup Tr^N(\Pi_1))\vdash \{p\} S \{q\}$. For $p\equiv true$ and $q\equiv false$, we have that $N\models \{true\} S \{false\}$; by Lemma \ref{alpha_defines_S}, it follows that $N\models \forall \vec{x},\vec{y} \neg \alpha_S(\vec{x},\vec{y})$; by Definition \ref{definition_of_pa_star}, it follows that $PA^* \vdash \forall \vec{x},\vec{y} \neg \alpha_S(\vec{x},\vec{y})$; then $PA^* \vdash p(\vec{x})\wedge \alpha_S(\vec{x},\vec{y})\rightarrow q(\vec{y}/\vec{x})$ follows; by Theorem \ref{from_hl_to_pa}, it follows that $HL(PA^*)\vdash \{p\} S \{q\}$.

We then show that for any $T\supseteq PA$ with $Thm(T)\subsetneqq Thm(PA^*)$, $HL(T)$ is not complete relative to $N$ for $\{Cnt\}WP\{Cnt\}$ with nonstandard inputs. By Definition \ref{definition_of_completeness_of_Hoare_logic} (i), we have to prove that for any $T\supseteq PA$ with $Thm(T)\subsetneqq Thm(PA^*)$, there exist $p,q\in Cnt$, and $S\in WP$ such that $N\models \{p\} S \{q\}$ but $HL(T)\not\vdash \{p\} S \{q\}$. Let $T\supseteq PA$ with $Thm(T)\subsetneqq Thm(PA^*)$. By Definition \ref{definition_of_pa_star}, it follows that there exists $S\in WP$ such that $N\models \forall \vec{x},\vec{y} \neg \alpha_S(\vec{x},\vec{y})$ and $T\not\vdash \forall \vec{x},\vec{y} \neg \alpha_S(\vec{x},\vec{y})$. Let $p ::= true$, $q ::= false$, and $S\in WP$ such that $N\models \forall \vec{x},\vec{y} \neg \alpha_S(\vec{x},\vec{y})$ and $T\not\vdash \forall \vec{x},\vec{y} \neg \alpha_S(\vec{x},\vec{y})$; by Lemma \ref{alpha_defines_S}, it follows that $N\models \{p\}S\{q\}$; since $T \vdash p(\vec{x})\wedge \alpha_S(\vec{x},\vec{y})\rightarrow q(\vec{y}/\vec{x})$, by Theorem \ref{from_hl_to_pa}, it follows that $HL(T)\not\vdash \{p\} S \{q\}$.
\end{proof}

\subsection{Hierarchy of $Th(N)$}\label{hierarchy of_Th_N}

\theoremstyle{plain}
\newtheorem{from_pi_to_sigma}{Lemma}[subsection]
\begin{from_pi_to_sigma}\label{from_pi_to_sigma}
  For any $n \geq 0$, $PA \cup Tr^N(\Pi_n) \vdash Tr^N(\Sigma_{n+1})$.
\end{from_pi_to_sigma}
\begin{proof}
Fix $n \geq 0$, and fix $\varphi\in Tr^N(\Sigma_{n+1})$. It remains to prove that $PA \cup Tr^N(\Pi_n) \vdash \varphi$. By definition of $\Sigma_{n+1}$, there exists a $\psi(x)\in\Pi_n$ such that $\varphi \equiv \exists x\ \psi(x)$. Since $N\models \varphi$, it follows that there exists $m\in N$ such that $N\models \psi(m)$. Since $\psi(m)$ is a $\Pi_n$-sentence, it follows that $PA \cup Tr^N(\Pi_n) \vdash \psi(m)$. By introducing the existential quantifier $\exists x$, it follows that $PA \cup Tr^N(\Pi_n) \vdash \exists x\ \psi(x)$. By definition of $\varphi$, we have that $PA \cup Tr^N(\Pi_n) \vdash \varphi$.
\end{proof}

\newtheorem{definability_of_sigma_and_pi}[from_pi_to_sigma]{Lemma}
\begin{definability_of_sigma_and_pi}\label{definability_of_sigma_and_pi}
  For any $n>0$, the sets of sentences $Tr^N(\Sigma_n)$, $Tr^N(\Pi_n)$, and $Thm(PA\cup Tr^N(\Pi_n))$ are $\Sigma_n$, $\Pi_n$, and $\Sigma_{n+1}$, respectively.
\end{definability_of_sigma_and_pi}
\begin{proof}
Let $n = k+1$ with $k\geq 0$. The argument of this lemma proceeds by induction on $k$.

We first prove that the lemma holds for $k = 0$. Consider $\varphi\in Tr^N(\Sigma_1)$ as follows: by definition of $Tr^N(\Sigma_1)$, it is equivalent to saying that $\varphi\in \Sigma_1$ and $N\models \varphi$; since a $\Sigma_1$-sentence is true in $N$ iff it is a theorem of $PA$, it is equivalent to saying that $\varphi\in \Sigma_1$ and $\varphi \in Thm(PA)$. Since $\Sigma_1$ is $\Delta_1$ and hence $\Sigma_1$, and $Thm(PA)$ is $\Sigma_1$, by the closure of $\Sigma_1$ under conjunction, it follows that $Tr^N(\Sigma_1)$ is $\Sigma_1$. Consider $\varphi\in Tr^N(\Pi_1)$ as follows: by definition of $Tr^N(\Pi_1)$, it is equivalent to saying that $\varphi\in \Pi_1$ and $N\models \varphi$; it is equivalent to saying that $\varphi\in \Pi_1$ and $N\not\models \neg\varphi$; since a $\Sigma_1$-sentence is true in $N$ iff it is a theorem of $PA$, and $\varphi\in\Pi_1$ iff $\neg\varphi\in\Sigma_1$, it is equivalent to saying that $\varphi\in \Pi_1$ and $\neg\varphi \not\in Thm(PA)$. It follows that $\Pi_1$ is $\Delta_1$ and hence $\Pi_1$. Since $Thm(PA)$ is $\Sigma_1$, we have that the set $\{\varphi : \neg\varphi \not\in Thm(PA)\}$ is $\Pi_1$. $\Pi_1$ being closed under conjunction, it follows that $Tr^N(\Pi_1)$ is $\Pi_1$. Since $PA$ is $\Delta_1$ and hence $\Sigma_2$, and $Tr^N(\Pi_1)$ is $\Pi_1$ and hence $\Sigma_2$, by the closure of $\Sigma_2$ under disjunction, it follows that $PA\cup Tr^N(\Pi_1)$ is $\Sigma_2$. By definition of $Thm(PA\cup Tr^N(\Pi_1))$, we remark that $Thm(PA\cup Tr^N(\Pi_1))$ is r.e. in $PA\cup Tr^N(\Pi_1)$ and hence $\Sigma_1$ in $PA\cup Tr^N(\Pi_1)$, so finally $\Sigma_2$.

Suppose that the lemma holds for $k\geq 0$, i.e., $Tr^N(\Sigma_{n-1})$, $Tr^N(\Pi_{n-1})$, and $Thm(PA\cup Tr^N(\Pi_{n-1}))$ are $\Sigma_{n-1}$, $\Pi_{n-1}$, and $\Sigma_n$, respectively. Then we have to prove that it also holds for $k+1$, i.e., $Tr^N(\Sigma_n)$, $Tr^N(\Pi_n)$, and $Thm(PA\cup Tr^N(\Pi_n))$ are $\Sigma_n$, $\Pi_n$, and $\Sigma_{n+1}$, respectively. By Lemma \ref{from_pi_to_sigma}, we have that $\varphi\in Tr^N(\Sigma_n)$ is equivalent to $\varphi\in \Sigma_n$ and $\varphi \in Thm(PA\cup Tr^N(\Pi_{n-1}))$. Since $\Sigma_n$ is $\Delta_1$ and hence $\Sigma_n$, and $Thm(PA\cup Tr^N(\Pi_{n-1}))$ is $\Sigma_n$, by the closure of $\Sigma_n$ under conjunction, it follows that $Tr^N(\Sigma_n)$ is $\Sigma_n$. Consider $\varphi\in Tr^N(\Pi_n)$ as follows: by definition of $Tr^N(\Pi_n)$, it is equivalent to saying that $\varphi\in \Pi_n$ and $N\models \varphi$; by pure logic, it is equivalent to saying that $\varphi\in \Pi_n$ and $N\not\models \neg\varphi$; since $\varphi\in\Pi_n$ iff $\neg\varphi\in\Sigma_n$, by Lemma \ref{from_pi_to_sigma}, it is equivalent to saying that $\varphi\in \Pi_n$ and $\neg\varphi \not\in Thm(PA\cup Tr^N(\Pi_{n-1}))$. For $\Pi_n$ is $\Delta_1$ and hence $\Pi_n$, and $\{\varphi : \neg\varphi \not\in Thm(PA\cup Tr^N(\Pi_{n-1}))\}$ is $\Pi_n$, by the closure of $\Pi_n$ under conjunction, it follows that $Tr^N(\Pi_n)$ is $\Pi_n$. Since $PA$ is $\Delta_1$ and hence $\Sigma_{n+1}$, and $Tr^N(\Pi_n)$ is $\Pi_n$ and hence $\Sigma_{n+1}$, by the closure of $\Sigma_{n+1}$ under disjunction, it follows that $PA\cup Tr^N(\Pi_n)$ is $\Sigma_{n+1}$. By definition of $Thm(PA\cup Tr^N(\Pi_n))$, we remark that $Thm(PA\cup Tr^N(\Pi_n))$ is r.e. in $PA\cup Tr^N(\Pi_n)$ and hence $\Sigma_1$ in $PA\cup Tr^N(\Pi_n)$, so finally $\Sigma_{n+1}$.
\end{proof}

\newtheorem{from_pi_to_pi}[from_pi_to_sigma]{Theorem}
\begin{from_pi_to_pi}\label{from_pi_to_pi}
  For any $n \geq 0$, $PA \cup Tr^N(\Pi_n) \not\vdash Tr^N(\Pi_{n+1})$.
\end{from_pi_to_pi}
\begin{proof}
The case for $n = 0$ follows from G\"{o}del's first completeness theorem, together with the fact that $PA \vdash Tr^N(\Pi_0)$. It remains to consider the cases for $n > 0$. Fix $n > 0$. By Lemma \ref{definability_of_sigma_and_pi}, $Thm(PA\cup Tr^N(\Pi_n))$ is $\Sigma_{n+1}$. Then there exists $\varphi(x)\in\Sigma_{n+1}$ such that for any $\psi\in L$,
\begin{equation}
  \psi\in Thm(PA\cup Tr^N(\Pi_n)) \mbox{ iff } N\models \varphi(\ulcorner\psi\urcorner). \tag{1}
\end{equation}
By G\"{o}del's diagonal lemma, there exists a sentence $G\in L$ such that
\begin{equation}
  PA \cup Tr^N(\Pi_n) \vdash G \leftrightarrow \neg \varphi(\ulcorner G\urcorner). \tag{2}
\end{equation}
Assume for a contradiction that $PA \cup Tr^N(\Pi_n) \vdash G$. Then $G\in Thm(PA\cup Tr^N(\Pi_n))$ and hence by assertion (1) we have $N\models \varphi(\ulcorner G\urcorner)$. On the other hand, by assertion (2), it follows that $PA \cup Tr^N(\Pi_n) \vdash \neg \varphi(\ulcorner G\urcorner)$. Since $N\models PA \cup Tr^N(\Pi_n)$, by soundness of first-order logic, we have that $N \models \neg \varphi(\ulcorner G\urcorner)$, contrary to $N\models \varphi(\ulcorner G\urcorner)$. So we have that $PA \cup Tr^N(\Pi_n) \not\vdash G$. Then $G\not\in Thm(PA\cup Tr^N(\Pi_n))$ follows. By assertion (1), it follows that $N\models \neg\varphi(\ulcorner G\urcorner)$. Since $\neg\varphi(\ulcorner G\urcorner)\in \Pi_{n+1}$, we have that $\neg\varphi(\ulcorner G\urcorner)\in Tr^N(\Pi_{n+1})$. By assertion (2), together with the fact $PA \cup Tr^N(\Pi_n) \not\vdash G$, it follows that $PA \cup Tr^N(\Pi_n) \not\vdash \neg\varphi(\ulcorner G\urcorner)$. Finally we have that $PA \cup Tr^N(\Pi_n) \not\vdash Tr^N(\Pi_{n+1})$.
\end{proof}

\subsection{When $P,Q$ $=$ $\Sigma_i, \Pi_i$, $i\geq 0$}\label{when_P_and_Q_range_over_the_arithmetical_hierarchy_with_nonstandard_inputs}

To investigate the completeness of $HL(T)$ relative to $N$ for $\{P\}WP\{Q\}$ with nonstandard inputs, we remark that if $P$ or $Q$ is too large, or $Thm(T)$ is too small, then $HL(T)$ might not be complete relative to $N$ for $\{P\}WP\{Q\}$ with nonstandard inputs. Hence we give that

\theoremstyle{definition}
\newtheorem{definitions_for_completeness_of_Hoare_logic}{Definition}[subsection]
\begin{definitions_for_completeness_of_Hoare_logic}\label{definitions_for_completeness_of_Hoare_logic}
  If $HL(T)$ is complete relative to $N$ for $\{P\}WP\{Q\}$ with nonstandard inputs, then we say that

  (i) pre-$P$ (resp. post-$Q$) is maximal w.r.t. $T$ with nonstandard inputs if for any $P'\not\subseteq P$ (resp. $Q'\not\subseteq Q$), $HL(T)$ is not complete relative to $N$ for $\{P'\}WP\{Q\}$ (resp. for $\{P\}WP\{Q'\}$) with nonstandard inputs.

  (ii) $T$ is minimal w.r.t. pre-$P$ (resp. w.r.t. post-$Q$) with nonstandard inputs if for any $T'\supseteq PA$ with $Thm(T')\subsetneqq Thm(T)$, $HL(T')$ is not complete relative to $N$ for $\{P\}AP\{Cnt\}$ (resp. for $\{Cnt\}AP\{Q\}$) with nonstandard inputs.
\end{definitions_for_completeness_of_Hoare_logic}

Note that in Definition \ref{definitions_for_completeness_of_Hoare_logic} (ii), in case $HL(T')$ is not complete relative to $N$ for $\{P\}AP\{Cnt\}$ (resp. for $\{Cnt\}AP\{Q\}$) with nonstandard inputs, we can see that $P$ (resp. $Q$) is the only factor leading to this, since $HL(PA)$ is complete relative to $N$ for $\{Cnt\}AP\{Cnt\}$ with nonstandard inputs (cf. Corollary \ref{completeness_of_Hoare_logic_for_Cnt_AP_Cnt}).

\theoremstyle{plain}
\newtheorem{completeness_of_Hoare_logic_for_sigma_i_WP_pi_i}[definitions_for_completeness_of_Hoare_logic]{Lemma}
\begin{completeness_of_Hoare_logic_for_sigma_i_WP_pi_i}\label{completeness_of_Hoare_logic_for_sigma_i_WP_pi_i}
  For any $i>0$, $HL(PA\cup Tr^N(\Pi_i))$ is complete relative to $N$ for $\{\Sigma_i\}WP\{\Pi_i\}$ with nonstandard inputs.
\end{completeness_of_Hoare_logic_for_sigma_i_WP_pi_i}
\begin{proof}
Fix $i>0$. Recalling Definition \ref{definition_of_completeness_of_Hoare_logic} (i), we have to prove that for any $p\in \Sigma_i$, $S\in WP$, and $q\in \Pi_i$, $N\models \{p\} S \{q\}$ implies $HL(PA\cup Tr^N(\Pi_i))\vdash \{p\} S \{q\}$. Let $N\models \{p\} S \{q\}$ with $S\in WP$ (having program variables $\vec{x}$), $p(\vec{u},\vec{x})\in \Sigma_i$ and $q(\vec{u},\vec{x})\in \Pi_i$. It remains to prove that $HL(PA\cup Tr^N(\Pi_i))\vdash \{p\} S \{q\}$. By Lemma \ref{alpha_defines_S}, it follows that $N\models \forall \vec{u},\vec{x}, \vec{y} (p(\vec{u},\vec{x})\wedge \alpha_S(\vec{x},\vec{y})\rightarrow q(\vec{u},\vec{y}/\vec{x}))$. By pure logic, we have that $N\models \forall \vec{u},\vec{x}, \vec{y} (\neg p(\vec{u},\vec{x})\vee \neg \alpha_S(\vec{x},\vec{y})\vee q(\vec{u},\vec{y}/\vec{x}))$. Since $p(\vec{u},\vec{x})$, $\alpha_S(\vec{x},\vec{y}) \in \Sigma_i$, it follows that $\neg p(\vec{u},\vec{x}), \neg\alpha_S(\vec{x},\vec{y}) \in \Pi_i$. By the closure of $\Pi_i$ under disjunction, it follows that $\neg p(\vec{u},\vec{x})\vee \neg \alpha_S(\vec{x},\vec{y})\vee q(\vec{u},\vec{y}/\vec{x})\in\Pi_i$. Then $\forall \vec{u},\vec{x}, \vec{y} (p(\vec{u},\vec{x})\wedge \alpha_S(\vec{x},\vec{y})\rightarrow q(\vec{u},\vec{y}/\vec{x}))\in Tr^N(\Pi_i)$ and hence $PA\cup Tr^N(\Pi_i)\vdash \forall \vec{u},\vec{x}, \vec{y} (p(\vec{u},\vec{x})\wedge \alpha_S(\vec{x},\vec{y})\rightarrow q(\vec{u},\vec{y}/\vec{x}))$. By Theorem \ref{from_hl_to_pa}, it follows that $HL(PA\cup Tr^N(\Pi_i))\vdash \{p\} S \{q\}$.
\end{proof}

\newtheorem{unprovability_of_a_class_of_Hoare_triples}[definitions_for_completeness_of_Hoare_logic]{Lemma}
\begin{unprovability_of_a_class_of_Hoare_triples}\label{unprovability_of_a_class_of_Hoare_triples}
Let $S ::= y:=0; while\ y<x\ do\ y:=y+1\ od$, and let $PA\subseteq T \subseteq Th(N)$, $\varphi(x)\in L$ such that $N\models \forall x\ \varphi(x)$ and $T\nvdash \forall x\ \varphi(x)$. It is the case that $HL(T)\nvdash \{\neg\varphi(x)\} S \{false\}$.
\end{unprovability_of_a_class_of_Hoare_triples}
\begin{proof}
  Follows from the proof of Theorem 4.3 of \cite{bergstra_2}.
\end{proof}

\newtheorem{maximality_of_assertions_w_r_t_theory_with_nonstandard_inputs}[definitions_for_completeness_of_Hoare_logic]{Lemma}
\begin{maximality_of_assertions_w_r_t_theory_with_nonstandard_inputs}\label{maximality_of_assertions_w_r_t_theory_with_nonstandard_inputs}
  Pre-$\Sigma_i$ (resp. post-$\Pi_i$) is maximal w.r.t. $PA \cup Tr^N(\Pi_i)$ with nonstandard inputs.
\end{maximality_of_assertions_w_r_t_theory_with_nonstandard_inputs}
\begin{proof}
Proof of pre-$\Sigma_i$ being maximal w.r.t. $PA \cup Tr^N(\Pi_i)$ with nonstandard inputs. Recalling Definition \ref{definitions_for_completeness_of_Hoare_logic} (i), we have to prove that there exist $p\in \Pi_i$ (the minimal level $\not\subseteq\Sigma_i$), $S\in WP$, and $q\in \Pi_i$ such that $N\models \{p\} S \{q\}$ but $HL(PA \cup Tr^N(\Pi_i))\not\vdash \{p\} S \{q\}$. By Theorem \ref{from_pi_to_pi}, it follows that $PA \cup Tr^N(\Pi_i)\not\vdash Tr^N(\Pi_{i+1})$. Then there exists a $\Pi_{i+1}$-sentence $\varphi$ such that $N\models \varphi$ and $PA \cup Tr^N(\Pi_i)\not\vdash \varphi$. By definition of $\Pi_{i+1}$, we have that, for some $\psi(x)\in \Sigma_i$, $\varphi \equiv \forall x\ \psi(x)$. Let $p ::= \neg\psi(x)$ ($\in \Pi_i$), $S ::= y:=0; while\ y<x\ do\ y:=y+1\ od$, and $q ::= false$. It's easy to check that $N\models \{p\} S \{q\}$. By Lemma \ref{unprovability_of_a_class_of_Hoare_triples}, it follows that $HL(PA \cup Tr^N(\Pi_i))\nvdash \{p\} S \{q\}$.

Proof of post-$\Pi_i$ being maximal w.r.t. $PA \cup Tr^N(\Pi_i)$ with nonstandard inputs. Recalling Definition \ref{definitions_for_completeness_of_Hoare_logic} (i), we have to prove that there exist $p\in \Sigma_i$, $S\in WP$, and $q\in \Sigma_i$ (the minimal level $\not\subseteq\Pi_i$) such that $N\models \{p\} S \{q\}$ but $HL(PA \cup Tr^N(\Pi_i))\not\vdash \{p\} S \{q\}$. Let $p\equiv true$, let $S ::= x := x$, and let $q\equiv \psi(x)$ with $\psi(x)$ being as defined in the proof of pre-$\Sigma_i$ being maximal w.r.t. $PA \cup Tr^N(\Pi_i)$ with nonstandard inputs. It's easy to see that $N\models \{p\}S\{q\}$. It remains to show that $HL(PA \cup Tr^N(\Pi_i))\not\vdash \{p\} S \{q\}$. By Theorem \ref{from_hl_to_pa}, it suffices to prove that $PA \cup Tr^N(\Pi_i)\not\vdash \forall x,y (true\wedge \alpha_S(x,y)\rightarrow \psi(y))$. By definition of $\alpha_S(x,y)$, it suffices to prove that $PA \cup Tr^N(\Pi_i)\not\vdash \forall x\ \psi(x)$. This is the case due to the choice of $\psi(x)$.
\end{proof}

By Lemma \ref{completeness_of_Hoare_logic_for_sigma_i_WP_pi_i}, together with Definition \ref{definition_of_completeness_of_Hoare_logic}, it follows that $HL(PA\cup Tr^N(\Pi_i))$ is complete relative to $N$ for $\{\Pi_{i-1}\}WP\{\Sigma_{i-1}\}$ with nonstandard inputs.

\newtheorem{minimality_of_theory_w_r_t_assertions_with_nonstandard_inputs}[definitions_for_completeness_of_Hoare_logic]{Lemma}
\begin{minimality_of_theory_w_r_t_assertions_with_nonstandard_inputs}\label{minimality_of_theory_w_r_t_assertions_with_nonstandard_inputs}
  $PA \cup Tr^N(\Pi_i)$ is minimal w.r.t. pre-$\Pi_{i-1}$ (resp. w.r.t. post-$\Sigma_{i-1}$) with nonstandard inputs.
\end{minimality_of_theory_w_r_t_assertions_with_nonstandard_inputs}
\begin{proof}
Proof of $PA \cup Tr^N(\Pi_i)$ being minimal w.r.t. pre-$\Pi_{i-1}$ with nonstandard inputs. Recalling Definition \ref{definitions_for_completeness_of_Hoare_logic} (ii), we have to prove that for any $T\supseteq PA$ with $Thm(T)\subsetneqq Thm(PA \cup Tr^N(\Pi_i))$, there exist $p\in \Pi_{i-1}$, $S\in AP$, and $q\in Cnt$ such that $N\models \{p\} S \{q\}$ but $HL(T)\not\vdash \{p\} S \{q\}$. Let $T\supseteq PA$ with $Thm(T)\subsetneqq Thm(PA \cup Tr^N(\Pi_i))$. Then there exists a $\Pi_i$-sentence $\varphi$ such that $N\models \varphi$ and $T\not\vdash \varphi$. By definition of $\Pi_i$, we have that, for some $\psi(x)\in \Sigma_{i-1}$, $\varphi \equiv \forall x\ \psi(x)$. Let $p ::= \neg \psi(x)$ ($\in \Pi_{i-1}$), $S ::= x:=x$, and $q ::= false$. It's easy to see that $N\models \{p\} S \{q\}$. It remains to show that $HL(T)\not\vdash \{p\} S \{q\}$. By Theorem \ref{from_hl_to_pa}, it suffices to prove that $T\not\vdash \forall x,y (\neg \psi(x) \wedge \alpha_S(x,y)\rightarrow false)$. Since $N\models \varphi$ and $T\not\vdash \varphi$, by completeness of first-order logic, there exists nonstandard $M\models T$ such that $M\models \exists x\ \neg\psi(x)$. Since $M\models \forall x \exists y\ \alpha_S(x,y)$, we have that $M \not\models \forall x,y (\neg \psi(x) \wedge \alpha_S(x,y)\rightarrow false)$. By completeness of first-order logic, it follows that $T\not\vdash \forall x,y (\neg \psi(x) \wedge \alpha_S(x,y)\rightarrow false)$.

Proof of $PA \cup Tr^N(\Pi_i)$ being minimal w.r.t. post-$\Sigma_{i-1}$ with nonstandard inputs. Recalling Definition \ref{definitions_for_completeness_of_Hoare_logic} (ii), we have to prove that for any $T\supseteq PA$ with $Thm(T)\subsetneqq Thm(PA \cup Tr^N(\Pi_i))$, there exist $p\in Cnt$, $S\in AP$, and $q\in \Sigma_{i-1}$ such that $N\models \{p\} S \{q\}$ but $HL(T)\not\vdash \{p\} S \{q\}$. Let $T\supseteq PA$ with $Thm(T)\subsetneqq Thm(PA \cup Tr^N(\Pi_i))$. Then there exists a $\Pi_i$-sentence $\varphi$ such that $N\models \varphi$ and $T\not\vdash \varphi$. By definition of $\Pi_i$, we have that, for some $\psi(x)\in \Sigma_{i-1}$, $\varphi \equiv \forall x\ \psi(x)$. Let $p ::= true$, $S ::= x := x$, and $q ::= \psi(x)$. It's easy to see that $N\models \{p\} S \{q\}$. It remains to show that $HL(T)\not\vdash \{p\} S \{q\}$. By Theorem \ref{from_hl_to_pa}, it suffices to prove that $T\not\vdash \forall x,y (true\wedge \alpha_S(x,y)\rightarrow \psi(y))$. Since $N\models \varphi$ and $T\not\vdash \varphi$, by completeness of first-order logic, there exists nonstandard $M\models T$ such that $M\models \exists x\ \neg\psi(x)$. Since $M\models \forall x\ \alpha_S(x,x)$, we have that $M \not\models \forall x,y (true\wedge \alpha_S(x,y)\rightarrow \psi(y))$. By completeness of first-order logic, it follows that $T\not\vdash \forall x,y (true\wedge \alpha_S(x,y)\rightarrow \psi(y))$.
\end{proof}

\newtheorem{completeness_of_Hoare_logic_with_nonstandard_inputs}[definitions_for_completeness_of_Hoare_logic]{Theorem}
\begin{completeness_of_Hoare_logic_with_nonstandard_inputs}\label{completeness_of_Hoare_logic_with_nonstandard_inputs}
For any $i>0$, it is the case that

  (i) $HL(PA\cup Tr^N(\Pi_i))$ is complete relative to $N$ for $\{P\}WP\{Q\}$ with nonstandard inputs iff $P\subseteq \Sigma_i$ and $Q\subseteq \Pi_i$;

  (ii) if $\Pi_{i-1} \subseteq P \subseteq \Sigma_i$ or $\Sigma_{i-1} \subseteq Q \subseteq \Pi_i$, then $HL(T)$ is complete relative to $N$ for $\{P\}WP\{Q\}$ with nonstandard inputs iff $Thm(T) \supseteq Thm(PA \cup Tr^N(\Pi_i))$.
\end{completeness_of_Hoare_logic_with_nonstandard_inputs}
\begin{proof}
By Definition \ref{definition_of_completeness_of_Hoare_logic} (i), together with Lemmas \ref{completeness_of_Hoare_logic_for_sigma_i_WP_pi_i}, \ref{maximality_of_assertions_w_r_t_theory_with_nonstandard_inputs} and \ref{minimality_of_theory_w_r_t_assertions_with_nonstandard_inputs}.
\end{proof}

\section{Completeness of $HL(T)$ for $\{P\}WP\{Q\}$ relative to $N$ without nonstandard inputs}

This section aims at studying the completeness of $HL(T)$ for $\{P\}WP\{Q\}$ relative to $N$ without nonstandard inputs: in Subsection \ref{when_P_and_Q_are_equal_to_Cnt_without_nonstandard_inputs}, the case when $P,Q$ $=$ $Cnt$ is investigated; in Subsection \ref{when_P_and_Q_range_over_the_arithmetical_hierarchy_without_nonstandard_inputs}, the case when $P,Q$ $=$ $\Sigma_i, \Pi_i$, $i\geq 0$, is investigated.

\subsection{When $P,Q$ $=$ $Cnt$}\label{when_P_and_Q_are_equal_to_Cnt_without_nonstandard_inputs}

\theoremstyle{plain}
\newtheorem{unprovability_of_the_halting_problem}{Lemma}[subsection]
\begin{unprovability_of_the_halting_problem}[{cf. \cite[Theorem 3.2.1]{xu16}}]\label{unprovability_of_the_halting_problem}
  There exist $S\in WP$ and $\vec{n}\in N$ such that $N\models \forall \vec{y} \neg \alpha_S(\vec{n},\vec{y})$ and $PA\nvdash \forall \vec{y} \neg \alpha_S(\vec{n},\vec{y})$.
\end{unprovability_of_the_halting_problem}

\theoremstyle{definition}
\newtheorem{definition_of_pa_plus}[unprovability_of_the_halting_problem]{Definition}
\begin{definition_of_pa_plus}[{The definition of $PA^+$, cf. \cite[Definition 3.2.2]{xu16}}]\label{definition_of_pa_plus}
  We define $PA^+$ to be
  \begin{eqnarray*}
    PA^+ &::=& PA \cup \{ \forall \vec{y} \neg \alpha_S(\vec{n},\vec{y}) : \vec{n}\in N \ \&\ S\in WP \\
     && \&\ N\models \forall \vec{y} \neg \alpha_S(\vec{n},\vec{y}) \ \&\ PA\nvdash \forall \vec{y} \neg \alpha_S(\vec{n},\vec{y})\}.
  \end{eqnarray*}
\end{definition_of_pa_plus}

\theoremstyle{plain}
\newtheorem{property_of_pa_plus}[unprovability_of_the_halting_problem]{Lemma}
\begin{property_of_pa_plus}\label{property_of_pa_plus}
It is the case that

(i) $Thm(PA)$ $\subsetneqq$ $Thm(PA^+)$ $\subseteq$ $Thm(PA \cup Tr^N(\Pi_1))$;

(ii) $PA^+$ and $Thm(PA^+)$ are $\Sigma_2$.
\end{property_of_pa_plus}
\begin{proof}
(i) $Thm(PA)$ $\subsetneqq$ $Thm(PA^+)$ follows from Lemma \ref{unprovability_of_the_halting_problem} and Definition \ref{definition_of_pa_plus}. $Thm(PA^+)$ $\subseteq$ $Thm(PA \cup Tr^N(\Pi_1))$ follows from Definition \ref{definition_of_pa_plus}, together with the fact that $\forall \vec{y} \neg \alpha_S(\vec{n},\vec{y})$ is logically equivalent to a $\Pi_1$-sentence.

(ii) Since $Thm(PA^+)$ is r.e. in $PA^+$, i.e. $\Sigma_1$ in $PA^+$, to prove $Thm(PA^+)$ is $\Sigma_2$, it suffices to prove that $PA^+$ is $\Sigma_2$. Consider the statement $\varphi\in PA^+$ as follows: by definition of $PA^+$, it is equivalent to saying that $\varphi\in PA$, or there exist $S\in WP$ and $\vec{n}\in N$ such that $\varphi = \forall \vec{y} \neg \alpha_S(\vec{n},\vec{y})$, $N\nvDash \neg\varphi$ and $PA\nvdash \varphi$; since $\neg\forall \vec{y} \neg \alpha_S(\vec{n},\vec{y})$ is logically equivalent to a $\Sigma_1$-sentence, and a $\Sigma_1$-sentence is true in $N$ iff it is a theorem of $PA$, it is equivalent to saying that $\varphi\in PA$, or there exist $S\in WP$ and $\vec{n}\in N$ such that $\varphi = \forall \vec{y} \neg \alpha_S(\vec{n},\vec{y})$, $\neg\varphi \not\in Thm(PA)$ and $\varphi\not\in Thm(PA)$. Note that the set $\{\varphi: \varphi = \forall \vec{y} \neg \alpha_S(\vec{n},\vec{y})\ \&\ S\in WP\ \&\ \vec{n}\in N\}$ is $\Delta_1$ and hence $\Sigma_2$. Since $Thm(PA)$ is $\Sigma_1$, we have that the set $\{\varphi: \varphi \not\in Thm(PA)\}$ is $\Pi_1$ and hence $\Sigma_2$, and the set $\{\varphi: \neg\varphi \not\in Thm(PA)\}$ is $\Pi_1$ and hence $\Sigma_2$. By closure of $\Sigma_2$ under conjunction, it follows that the set $\{ \varphi: \varphi = \forall \vec{y} \neg \alpha_S(\vec{n},\vec{y})\ \&\ S\in WP\ \&\ \vec{n}\in N\ \&\ \neg\varphi \not\in Thm(PA)\ \&\ \varphi\not\in Thm(PA)\}$ is $\Sigma_2$. Moreover, since $PA$ is $\Delta_1$, we immediately have that the set $\{\varphi: \varphi\in PA\}$ is $\Sigma_2$. By closure of $\Sigma_2$ under disjunction, it follows that $PA^+$ is $\Sigma_2$.
\end{proof}

\theoremstyle{plain}
\newtheorem{completeness_of_Hoare_logic_for_Cnt_WP_Cnt_without_nonstandard_inputs}[unprovability_of_the_halting_problem]{Theorem}
\begin{completeness_of_Hoare_logic_for_Cnt_WP_Cnt_without_nonstandard_inputs}\label{completeness_of_Hoare_logic_for_Cnt_WP_Cnt_without_nonstandard_inputs}
  $PA^+$ is the minimal extension $T$ of $PA$ such that $HL(T)$ is complete relative to $N$ for $\{Cnt\}WP\{Cnt\}$ without nonstandard inputs.
\end{completeness_of_Hoare_logic_for_Cnt_WP_Cnt_without_nonstandard_inputs}
\begin{proof}
We first show that $HL(PA^+)$ is complete relative to $N$ for $\{Cnt\}WP\{Cnt\}$ without nonstandard inputs. By Definition \ref{definition_of_completeness_of_Hoare_logic} (ii), we have to prove that for any $S\in WP$ with program variables $\vec{x}$, $p,q\in Cnt$, and $\vec{n}\in N$, $N\models \{p \wedge \vec{x} = \vec{n}\} S \{q\}$ implies $HL(PA^+)\vdash \{p \wedge \vec{x} = \vec{n}\} S \{q\}$. Let $N\models \{p \wedge \vec{x} = \vec{n}\} S \{q\}$ with $p,q\in Cnt$, $S\in WP$ (having program variables $\vec{x}$), and $\vec{n}\in N$. It remains to prove that $HL(PA^+)\vdash \{p \wedge \vec{x} = \vec{n}\} S \{q\}$. For $p\equiv false$ or $q \equiv true$, it's easy to see that $PA^+ \vdash p(\vec{x}) \wedge \vec{x} = \vec{n} \wedge \alpha_S(\vec{x},\vec{y})\rightarrow q(\vec{y}/\vec{x})$; by Theorem \ref{from_hl_to_pa}, it follows that $HL(PA^+)\vdash \{p \wedge \vec{x} = \vec{n}\} S \{q\}$. For $p\equiv true$ and $q\equiv false$, we have that $N\models \{true \wedge \vec{x} = \vec{n}\} S \{false\}$; by Lemma \ref{alpha_defines_S}, it follows that $N\models \forall \vec{y} \neg \alpha_S(\vec{n},\vec{y})$; by Definition \ref{definition_of_pa_plus}, it follows that $PA^+ \vdash \forall \vec{y} \neg \alpha_S(\vec{n},\vec{y})$; then $PA^+ \vdash p(\vec{x}) \wedge \vec{x} = \vec{n} \wedge \alpha_S(\vec{x},\vec{y})\rightarrow q(\vec{y}/\vec{x})$ follows; by Theorem \ref{from_hl_to_pa}, it follows that $HL(PA^+)\vdash \{p \wedge \vec{x} = \vec{n}\} S \{q\}$.

We then show that for any $T\supseteq PA$ with $Thm(T)\subsetneqq Thm(PA^+)$, $HL(T)$ is not complete relative to $N$ for $\{Cnt\}WP\{Cnt\}$ without nonstandard inputs. By Definition \ref{definition_of_completeness_of_Hoare_logic} (ii), we have to prove that for any $T\supseteq PA$ with $Thm(T)\subsetneqq Thm(PA^+)$, there exist $S\in WP$ with program variables $\vec{x}$, $p,q\in Cnt$, and $\vec{n}\in N$ such that $N\models \{p \wedge \vec{x} = \vec{n}\} S \{q\}$ but $HL(T)\not\vdash \{p \wedge \vec{x} = \vec{n}\} S \{q\}$. Let $T\supseteq PA$ with $Thm(T)\subsetneqq Thm(PA^+)$. By Definition \ref{definition_of_pa_plus}, it follows that there exists $S\in WP$ and $\vec{n}\in N$ such that $N\models \forall \vec{y} \neg \alpha_S(\vec{n},\vec{y})$ and $T\not\vdash \forall \vec{y} \neg \alpha_S(\vec{n},\vec{y})$. Let $p ::= true$, $q ::= false$, and $S\in WP$, $\vec{n}\in N$ such that $N\models \forall \vec{y} \neg \alpha_S(\vec{n},\vec{y})$ and $T\not\vdash \forall \vec{y} \neg \alpha_S(\vec{n},\vec{y})$; by Lemma \ref{alpha_defines_S}, it follows that $N\models \{p \wedge \vec{x} = \vec{n}\}S\{q\}$; since $T \not\vdash p(\vec{x})\wedge \vec{x} = \vec{n} \wedge \alpha_S(\vec{x},\vec{y})\rightarrow q(\vec{y}/\vec{x})$, by Theorem \ref{from_hl_to_pa}, it follows that $HL(T)\not\vdash \{p \wedge \vec{x} = \vec{n}\} S \{q\}$.
\end{proof}

\subsection{When $P,Q$ $=$ $\Sigma_i, \Pi_i$, $i\geq 0$}\label{when_P_and_Q_range_over_the_arithmetical_hierarchy_without_nonstandard_inputs}

Similar to Definition \ref{definitions_for_completeness_of_Hoare_logic}, we give that

\theoremstyle{definition}
\newtheorem{definitions_for_completeness_of_Hoare_logic_without_nonstandard_inputs}{Definition}[subsection]
\begin{definitions_for_completeness_of_Hoare_logic_without_nonstandard_inputs}\label{definitions_for_completeness_of_Hoare_logic_without_nonstandard_inputs}
  If $HL(T)$ is complete relative to $N$ for $\{P\}WP\{Q\}$ without nonstandard inputs, then we say that

  (i) pre-$P$ (resp. post-$Q$) is maximal w.r.t. $T$ without nonstandard inputs if for any $P'\not\subseteq P$ (resp. $Q'\not\subseteq Q$), $HL(T)$ is not complete relative to $N$ for $\{P'\}WP\{Q\}$ (resp. for $\{P\}WP\{Q'\}$) without nonstandard inputs.

  (ii) $T$ is minimal w.r.t. pre-$P$ (resp. w.r.t. post-$Q$) without nonstandard inputs if for any $T'\supseteq PA$ with $Thm(T')\subsetneqq Thm(T)$, $HL(T')$ is not complete relative to $N$ for $\{P\}AP\{Cnt\}$ (resp. for $\{Cnt\}AP\{Q\}$) without nonstandard inputs.
\end{definitions_for_completeness_of_Hoare_logic_without_nonstandard_inputs}

\newtheorem{definability_of_recursive_functions}[definitions_for_completeness_of_Hoare_logic_without_nonstandard_inputs]{Lemma}
\begin{definability_of_recursive_functions}[{cf. \cite[Theorem 3.2.5]{xu16}}]\label{definability_of_recursive_functions}
For every $S\in WP$, every $M\models PA^+$ and every $\vec{n}\in N$, $f_S^M(\vec{n}) = \vec{y}$ iff $M\models \alpha_S(\vec{n},\vec{y})$.
\end{definability_of_recursive_functions}

\theoremstyle{plain}
\newtheorem{completeness_of_Hoare_logic_for_pi_i_plus_1_WP_sigma_i_plus_1}[definitions_for_completeness_of_Hoare_logic_without_nonstandard_inputs]{Lemma}
\begin{completeness_of_Hoare_logic_for_pi_i_plus_1_WP_sigma_i_plus_1}\label{completeness_of_Hoare_logic_for_pi_i_plus_1_WP_sigma_i_plus_1}
  For any $i>0$, $HL(PA\cup Tr^N(\Pi_i))$ is complete relative to $N$ for $\{\Pi_{i+1}\}WP\{\Sigma_{i+1}\}$ without nonstandard inputs.
\end{completeness_of_Hoare_logic_for_pi_i_plus_1_WP_sigma_i_plus_1}
\begin{proof}
Fix $i>0$. Recalling Definition \ref{definition_of_completeness_of_Hoare_logic} (ii), we have to prove that for any $S\in WP$ with program variables $\vec{x}$, $p(\vec{u},\vec{x})\in \Pi_{i+1}$, $q(\vec{u},\vec{x})\in \Sigma_{i+1}$ (besides $\vec{x}$, $p$ and $q$ could contain other free variables $\vec{u}$), and $\vec{m},\vec{n}\in N$, $N\models \{p \wedge (\vec{u},\vec{x}) = (\vec{m},\vec{n})\} S \{q\}$ implies $HL(PA\cup Tr^N(\Pi_i))\vdash \{p \wedge (\vec{u},\vec{x}) = (\vec{m},\vec{n})\} S \{q\}$. Let $N\models \{p \wedge (\vec{u},\vec{x}) = (\vec{m},\vec{n})\} S \{q\}$ with $S\in WP$ (having program variables $\vec{x}$), $p(\vec{u},\vec{x})\in \Pi_{i+1}$, $q(\vec{u},\vec{x})\in \Sigma_{i+1}$ and $\vec{m},\vec{n}\in N$. It remains to prove that $HL(PA\cup Tr^N(\Pi_i))\vdash \{p \wedge (\vec{u},\vec{x}) = (\vec{m},\vec{n})\} S \{q\}$. By Lemma \ref{alpha_defines_S}, it follows that $N\models \forall \vec{y} ( p(\vec{m},\vec{n}) \wedge \alpha_S(\vec{n},\vec{y}) \rightarrow q(\vec{m},\vec{y}) )$. To prove $HL(PA\cup Tr^N(\Pi_i))\vdash \{p \wedge (\vec{u},\vec{x}) = (\vec{m},\vec{n})\} S \{q\}$, by Theorem \ref{from_hl_to_pa}, it suffices to prove that $PA\cup Tr^N(\Pi_i)\vdash \forall \vec{y} ( p(\vec{m},\vec{n}) \wedge \alpha_S(\vec{n},\vec{y}) \rightarrow q(\vec{m},\vec{y}) )$. Let $M\models PA\cup Tr^N(\Pi_i)$ be arbitrary but fixed. By completeness of first-order logic, it suffices to prove that $M\models \forall \vec{y} ( p(\vec{m},\vec{n}) \wedge \alpha_S(\vec{n},\vec{y}) \rightarrow q(\vec{m},\vec{y}) )$. Suppose, for some $\vec{y}\in M$, that $M\models p(\vec{m},\vec{n}) \wedge \alpha_S(\vec{n},\vec{y})$. Then we have to prove that $M\models q(\vec{m},\vec{y})$. Since $p(\vec{m},\vec{n})$ is a $\Pi_{i+1}$-sentence, we have that, for some $\varphi(x)\in \Sigma_i$, $p(\vec{m},\vec{n}) \equiv \forall x\ \varphi(x)$. Assume that $N\not\models \forall x\ \varphi(x)$. Then, for some $\vec{r}\in N$, $N\models \neg\varphi(\vec{r})$. Since $\neg\varphi(\vec{r})$ is logically equivalent to a true $\Pi_i$-sentence, it follows that $PA\cup Tr^N(\Pi_i)\vdash \neg\varphi(\vec{r})$. By soundness of first-order logic, we have that $M\models \neg\varphi(\vec{r})$, contradicting the supposition $M\models \forall x\ \varphi(x)$. Then $N\models p(\vec{m},\vec{n})$ follows. By Lemma \ref{property_of_pa_plus} (i), it follows that $M\models PA^+$. By Lemmas \ref{definability_of_recursive_functions} and \ref{alpha_defines_S}, there exists $\vec{s}\in N$ such that $\vec{y} = \vec{s}$ and $N\models \alpha_S(\vec{n},\vec{s})$. Since $N\models p(\vec{m},\vec{n})\wedge \alpha_S(\vec{n},\vec{s})$, it follows from $N\models \forall \vec{y} ( p(\vec{m},\vec{n}) \wedge \alpha_S(\vec{n},\vec{y}) \rightarrow q(\vec{m},\vec{y}) )$ that $N\models q(\vec{m},\vec{s})$. Since $q(\vec{m},\vec{s})$ is a true $\Sigma_{i+1}$-sentence, by Lemma \ref{from_pi_to_sigma}, we have that $PA\cup Tr^N(\Pi_i)\vdash q(\vec{m},\vec{s})$. Then by soundness of first-order logic $M\models q(\vec{m},\vec{s})$ follows, and finally we have that $M\models q(\vec{m},\vec{y})$.
\end{proof}

\newtheorem{maximality_of_assertions_w_r_t_theory_without_nonstandard_inputs}[definitions_for_completeness_of_Hoare_logic_without_nonstandard_inputs]{Lemma}
\begin{maximality_of_assertions_w_r_t_theory_without_nonstandard_inputs}\label{maximality_of_assertions_w_r_t_theory_without_nonstandard_inputs}
  Pre-$\Pi_{i+1}$ (resp. post-$\Sigma_{i+1}$) is maximal w.r.t. $PA \cup Tr^N(\Pi_i)$ without nonstandard inputs.
\end{maximality_of_assertions_w_r_t_theory_without_nonstandard_inputs}
\begin{proof}
Proof of pre-$\Pi_{i+1}$ being maximal w.r.t. $PA \cup Tr^N(\Pi_i)$ without nonstandard inputs. Recalling Definition \ref{definitions_for_completeness_of_Hoare_logic_without_nonstandard_inputs} (i), we have to prove that there exist $p(\vec{u},\vec{x})\in \Sigma_{i+1}$ (the minimal level $\not\subseteq\Pi_{i+1}$), $S\in WP$, $q(\vec{u},\vec{x})\in \Sigma_{i+1}$, and $\vec{m},\vec{n}\in N$ such that $N\models \{p \wedge (\vec{u},\vec{x}) = (\vec{m},\vec{n})\} S \{q\}$ but $HL(PA\cup Tr^N(\Pi_i))\not\vdash \{p \wedge (\vec{u},\vec{x}) = (\vec{m},\vec{n})\} S \{q\}$. By Theorem \ref{from_pi_to_pi}, it follows that $PA \cup Tr^N(\Pi_i)\not\vdash Tr^N(\Pi_{i+1})$. Then there exists a $\Pi_{i+1}$-sentence $\varphi$ such that $N\models \varphi$ and $PA \cup Tr^N(\Pi_i)\not\vdash \varphi$. Let $p ::= \neg \varphi$ ($\in \Sigma_{i+1}$), $S ::= x:=x$, $q ::= false$, and $n\in N$. It's easy to see that $N\models \{p \wedge x = n \} S \{q\}$ (note that $\vec{u} = \emptyset$). It remains to show that $HL(PA \cup Tr^N(\Pi_i))\not\vdash \{p \wedge x = n \} S \{q\}$. By Theorem \ref{from_hl_to_pa}, it suffices to prove that $PA \cup Tr^N(\Pi_i)\not\vdash \forall y (\neg \varphi \wedge \alpha_S(n,y)\rightarrow false)$. By pure logic, it suffices to prove that $PA \cup Tr^N(\Pi_i)\not\vdash \varphi \vee \forall y\ \neg\alpha_S(n,y)$. This is the case due to the fact that $PA \cup Tr^N(\Pi_i)\not\vdash \varphi$ and $PA \cup Tr^N(\Pi_i)\vdash \exists y\ \alpha_S(n,y)$.

Proof of post-$\Sigma_{i+1}$ being maximal w.r.t. $PA \cup Tr^N(\Pi_i)$ without nonstandard inputs. Recalling Definition \ref{definitions_for_completeness_of_Hoare_logic_without_nonstandard_inputs} (i), we have to prove that there exist $p(\vec{u},\vec{x})\in \Pi_{i+1}$, $S\in WP$, $q(\vec{u},\vec{x})\in \Pi_{i+1}$ (the minimal level $\not\subseteq\Sigma_{i+1}$), and $\vec{m},\vec{n}\in N$ such that $N\models \{p \wedge (\vec{u},\vec{x}) = (\vec{m},\vec{n})\} S \{q\}$ but $HL(PA\cup Tr^N(\Pi_i))\not\vdash \{p \wedge (\vec{u},\vec{x}) = (\vec{m},\vec{n})\} S \{q\}$. Let $p\equiv true$, $S ::= x := x$, $q\equiv \varphi$ with $\varphi$ being as defined in the proof of pre-$\Pi_{i+1}$ being maximal w.r.t. $PA \cup Tr^N(\Pi_i)$ without nonstandard inputs, and $n\in N$. It's easy to see that $N\models \{p \wedge x = n \} S \{q\}$. It remains to show that $HL(PA \cup Tr^N(\Pi_i))\not\vdash \{p \wedge x = n \} S \{q\}$. By Theorem \ref{from_hl_to_pa}, it suffices to prove that $PA \cup Tr^N(\Pi_i)\not\vdash \forall y (true\wedge \alpha_S(n,y)\rightarrow \varphi)$. By pure logic, it suffices to prove that $PA \cup Tr^N(\Pi_i)\not\vdash \forall y\ \neg\alpha_S(n,y)\vee \varphi$. This is the case due to the fact that $PA \cup Tr^N(\Pi_i)\vdash \exists y\ \alpha_S(n,y)$ and $PA \cup Tr^N(\Pi_i)\not\vdash \varphi$.
\end{proof}

By Lemma \ref{completeness_of_Hoare_logic_for_pi_i_plus_1_WP_sigma_i_plus_1}, together with Definition \ref{definition_of_completeness_of_Hoare_logic} (ii), it follows that $HL(PA\cup Tr^N(\Pi_i))$ is complete relative to $N$ for $\{\Sigma_i\}WP\{\Pi_i\}$ without nonstandard inputs.

\newtheorem{minimality_of_theory_w_r_t_assertions_without_nonstandard_inputs}[definitions_for_completeness_of_Hoare_logic_without_nonstandard_inputs]{Lemma}
\begin{minimality_of_theory_w_r_t_assertions_without_nonstandard_inputs}\label{minimality_of_theory_w_r_t_assertions_without_nonstandard_inputs}
  $PA \cup Tr^N(\Pi_i)$ is minimal w.r.t. pre-$\Sigma_i$ (resp. w.r.t. post-$\Pi_i$) without nonstandard inputs.
\end{minimality_of_theory_w_r_t_assertions_without_nonstandard_inputs}
\begin{proof}
Proof of $PA \cup Tr^N(\Pi_i)$ being minimal w.r.t. pre-$\Sigma_i$ without nonstandard inputs. Recalling Definition \ref{definitions_for_completeness_of_Hoare_logic_without_nonstandard_inputs} (ii), we have to prove that for any $T\supseteq PA$ with $Thm(T)\subsetneqq Thm(PA \cup Tr^N(\Pi_i))$, there exist $p(\vec{u},\vec{x})\in \Sigma_i$, $S\in AP$, $q(\vec{u},\vec{x})\in Cnt$, and $\vec{m},\vec{n}\in N$ such that $N\models \{p \wedge (\vec{u},\vec{x}) = (\vec{m},\vec{n})\} S \{q\}$ but $HL(T)\not\vdash \{p \wedge (\vec{u},\vec{x}) = (\vec{m},\vec{n})\} S \{q\}$. Let $T\supseteq PA$ with $Thm(T)\subsetneqq Thm(PA \cup Tr^N(\Pi_i))$. Then there exists a $\Pi_i$-sentence $\varphi$ such that $N\models \varphi$ and $T\not\vdash \varphi$. Let $p ::= \neg \varphi$ ($\in \Sigma_i$), $S ::= x:=x$, $q ::= false$, and $n\in N$. The proof of $N\models \{p \wedge x = n\} S \{q\}$ and $HL(T)\not\vdash \{p \wedge x = n\} S \{q\}$ is similar to the case of pre-$\Pi_{i+1}$ being maximal w.r.t. $PA \cup Tr^N(\Pi_i)$ without nonstandard inputs.

Proof of $PA \cup Tr^N(\Pi_i)$ is minimal w.r.t. post-$\Pi_i$ without nonstandard inputs. Recalling Definition \ref{definitions_for_completeness_of_Hoare_logic_without_nonstandard_inputs} (ii), we have to prove that for any $T\supseteq PA$ with $Thm(T)\subsetneqq Thm(PA \cup Tr^N(\Pi_i))$, there exist $p(\vec{u},\vec{x})\in Cnt$, $S\in AP$, $q(\vec{u},\vec{x})\in \Pi_i$, and $\vec{m},\vec{n}\in N$ such that $N\models \{p \wedge (\vec{u},\vec{x}) = (\vec{m},\vec{n})\} S \{q\}$ but $HL(T)\not\vdash \{p \wedge (\vec{u},\vec{x}) = (\vec{m},\vec{n})\} S \{q\}$. Let $T\supseteq PA$ with $Thm(T)\subsetneqq Thm(PA \cup Tr^N(\Pi_i))$. Then there exists a $\Pi_i$-sentence $\varphi$ such that $N\models \varphi$ and $T\not\vdash \varphi$. Let $p ::= true$, $S ::= x := x$, $q ::= \varphi$, and $n\in N$. The proof of $N\models \{p \wedge x = n\} S \{q\}$ and $HL(T)\not\vdash \{p \wedge x = n\} S \{q\}$ is similar to the case of post-$\Sigma_{i+1}$ being maximal w.r.t. $PA \cup Tr^N(\Pi_i)$ without nonstandard inputs.
\end{proof}

\newtheorem{completeness_of_Hoare_logic_without_nonstandard_inputs}[definitions_for_completeness_of_Hoare_logic_without_nonstandard_inputs]{Theorem}
\begin{completeness_of_Hoare_logic_without_nonstandard_inputs}\label{completeness_of_Hoare_logic_without_nonstandard_inputs}
For any $i>0$, it is the case that

  (i) $HL(PA\cup Tr^N(\Pi_i))$ is complete relative to $N$ for $\{P\}WP\{Q\}$ without nonstandard inputs iff $P\subseteq \Pi_{i+1}$ and $Q\subseteq \Sigma_{i+1}$;

  (ii) if $\Sigma_i \subseteq P \subseteq \Pi_{i+1}$ or $\Pi_i \subseteq Q \subseteq \Sigma_{i+1}$, then $HL(T)$ is complete relative to $N$ for $\{P\}WP\{Q\}$ without nonstandard inputs iff $Thm(T) \supseteq Thm(PA \cup Tr^N(\Pi_i))$.
\end{completeness_of_Hoare_logic_without_nonstandard_inputs}
\begin{proof}
By Definition \ref{definition_of_completeness_of_Hoare_logic} (ii), together with Lemmas \ref{completeness_of_Hoare_logic_for_pi_i_plus_1_WP_sigma_i_plus_1}, \ref{maximality_of_assertions_w_r_t_theory_without_nonstandard_inputs} and \ref{minimality_of_theory_w_r_t_assertions_without_nonstandard_inputs}.
\end{proof}

\section{Comparison of $PA^*$, $PA^+$ and $PA \cup Tr^N(\Pi_1)$}\label{x_recursion_theory}

Theorem \ref{completeness_of_Hoare_logic_for_Cnt_WP_Cnt_with_nonstandard_inputs} (resp. Theorem \ref{completeness_of_Hoare_logic_for_Cnt_WP_Cnt_without_nonstandard_inputs}) says that $PA^*$ (resp. $PA^+$) is the minimal extension $T$ of $PA$ such that $HL(T)$ is complete relative to $N$ for $\{Cnt\}WP\{Cnt\}$ with (resp. without) nonstandard inputs. To see the real effects of excluding nonstandard inputs on the completeness of $HL(T)$ relative to $N$ for $\{Cnt\}WP\{Cnt\}$, we need to compare $PA^*$ with $PA^+$.

By letting $\Pi_0\subseteq P \subseteq \Sigma_1$ or $\Sigma_0 \subseteq Q \subseteq \Pi_1$ (resp. $\Sigma_1\subseteq P \subseteq \Pi_2$ or $\Pi_1 \subseteq Q \subseteq \Sigma_2$), it follows from Theorem \ref{completeness_of_Hoare_logic_with_nonstandard_inputs} (resp. Theorem \ref{completeness_of_Hoare_logic_without_nonstandard_inputs}) that $PA\cup Tr^N(\Pi_1)$ is the minimal extension $T$ of $PA$ such that $HL(T)$ is complete relative to $N$ for $\{P\}WP\{Q\}$ with (resp. without) nonstandard inputs. Recalling Lemma \ref{minimality_of_theory_w_r_t_assertions_with_nonstandard_inputs} (resp. Lemma \ref{minimality_of_theory_w_r_t_assertions_without_nonstandard_inputs}), we find that the minimality of $PA\cup Tr^N(\Pi_1)$ in the above sense is due to the choices of $P$ and $Q$, and not determined by the complexity of $WP$ in logic (at least not explicitly stated). On the other hand, the minimality of $PA^*$ (resp. $PA^+$) in Theorem \ref{completeness_of_Hoare_logic_for_Cnt_WP_Cnt_with_nonstandard_inputs} (resp. Theorem \ref{completeness_of_Hoare_logic_for_Cnt_WP_Cnt_without_nonstandard_inputs}) is determined totally by the complexity of $WP$ in logic. To see the essential role of $WP$ in the minimality of $PA\cup Tr^N(\Pi_1)$ in the above sense, we need to compare $PA^*$ (resp. $PA^+$) with $PA\cup Tr^N(\Pi_1)$.

This section devotes to investigating the relationship of $PA^*$, $PA^+$ and $PA\cup Tr^N(\Pi_1)$. It will be established that $Thm(PA^*)$ $=$ $Thm(PA^+)$ $=$ $Thm(PA \cup Tr^N(\Pi_1))$. By Lemmas \ref{property_of_pa_star} and \ref{property_of_pa_plus}, it follows that $Thm(PA^*)$, $Thm(PA^+)$ $\subseteq$ $Thm(PA \cup Tr^N(\Pi_1))$. It remains to prove that $Thm(PA^*)$, $Thm(PA^+)$ $\supseteq$ $Thm(PA \cup Tr^N(\Pi_1))$. This technical line requires that the classical recursive functions, defined in $N$, be redefined in $PA$, called X-recursive functions, and, correspondingly, recursion theory be extended to X-recursion theory. The rest of this section is organized as follows: definition of X-recursive functions is given in Subsection \ref{definition_of_X_recursive_functions}; X-recursion theory is partly developed in Subsection \ref{properties_of_X_recursive_functions}; relationship of $PA^*$, $PA^+$ and $PA \cup Tr^N(\Pi_1)$ is established in Subsection \ref{relationship_of_PA_star_with_PA_plus}.

\subsection{The definition of X-recursive functions}\label{definition_of_X_recursive_functions}

Before defining X-recursive functions, the processes of composition, recursion and minimization are defined in $PA$ as follows.

\theoremstyle{definition}
\newtheorem{composition}{Definition}[section]
\begin{composition}[Composition]
Let $\varphi$ be an $m$-place $L$-formula such that $PA \vdash \forall \vec{x},y,y' (\varphi(\vec{x},y)\wedge \varphi(\vec{x},y')\rightarrow y=y')$, and $\psi_1\ldots \psi_m$ be $n$-place $L$-formulas such that for each $1\leq i \leq m$, $PA \vdash \forall \vec{x},y,y' (\psi_i(\vec{x},y)\wedge \psi_i(\vec{x},y')\rightarrow y=y')$, $\varphi$ and $\psi_1,\ldots,\psi_m$ defining functions $f$ and $g_1,\ldots,g_m$ respectively. Define $y=h(\vec{x})$ from $f$ and $g_1,\ldots,g_m$ by the $L$-formula $\theta(\vec{x},y)$, i.e. $\exists \vec{z} (\bigwedge_{1\leq i \leq m}\psi_i(\vec{x},z_i) \wedge \varphi(\vec{z},y))$. The process defined by $\theta$ from $f$ and $g_1,\ldots,g_m$ is called composition.
\end{composition}
\theoremstyle{plain}
\newtheorem{well_defined_composition}[composition]{Lemma}
\begin{well_defined_composition}
Let $f$, $g_1,\ldots,g_m$ and $h$ be as defined before. Then $PA$ proves that

(a) $\forall \vec{x},y ( \exists \vec{z}(\bigwedge_{1\leq i \leq m} g_i(\vec{x})=z_i \wedge f(\vec{z})=y) \leftrightarrow h(\vec{x})=y )$;

(b) $\forall \vec{x},y,y' ( h(\vec{x})=y \wedge h(\vec{x})=y' \rightarrow y=y' )$.
\end{well_defined_composition}

For convenience, we write $h$, the function obtained by composition from $f$ and $g_1,\ldots,g_m$, in the form: $h(\vec{x}) = f(g_1(\vec{x}),\ldots,g_m(\vec{x}))$.

\theoremstyle{definition}
\newtheorem{recursion}[composition]{Definition}
\begin{recursion}[Recursion]
Let $\varphi(\vec{x},y), \psi(\vec{x},y,z,w)\in L$ such that $PA \vdash \forall \vec{x}, y, y'(\varphi(\vec{x},y)\wedge \varphi(\vec{x},y')\rightarrow y=y')$ and $PA \vdash \forall \vec{x},y,z,w,w'(\psi(\vec{x},y,z,w)\wedge \psi(\vec{x},$ $y,z,w') \rightarrow w=w')$, $\varphi$ and $\psi$ defining functions $y = f(\vec{x})$ and $w = g(\vec{x},y,z)$ respectively. Define $z=h(\vec{x},y)$ from $f$ and $g$ by the $L$-formula $\theta(\vec{x},y,z)$, i.e. $\exists w( \varphi(\vec{x},(w)_0)\wedge \forall i<y\ \psi(\vec{x},i,(w)_i,(w)_{i+1} ) \wedge (w)_y=z )$. The process defined by $\theta$ from $f$ and $g$ is called (primitive) recursion.
\end{recursion}
\theoremstyle{plain}
\newtheorem{well_defined_recursion}[composition]{Lemma}
\begin{well_defined_recursion}\label{well_defined_recursion}
Let $f$, $g$ and $h$ be as defined before. Then $PA$ proves that

(a) $\forall \vec{x},z  (f(\vec{x})=z \leftrightarrow h(\vec{x},0) = z )$;

(b) $\forall \vec{x},y,z,z'(h(\vec{x},y)=z\wedge g(\vec{x},y,z)=z' \leftrightarrow h(\vec{x},y+1) = z' )$;

(c) $\forall \vec{x},y,z,z'(h(\vec{x},y)=z\wedge h(\vec{x},y)=z'\rightarrow z=z')$.
\end{well_defined_recursion}
\begin{proof}
Let $M\models PA$.

(a) Fix $\vec{a},b\in M$. To prove $PA\vdash \forall \vec{x},z  (f(\vec{x})=z \leftrightarrow h(\vec{x},0) = z )$, by completeness of first order logic, it suffices to prove that $M\models f(\vec{a})=b \leftrightarrow h(\vec{a},0) = b )$. Consider $M \models f(\vec{a})=b$ as follows: by Lemma \ref{beta_function} (a), it is equivalent to $M\models \exists w( f(\vec{a})=(w)_0 \wedge (w)_0=b )$; by definition of $h$, it is equivalent to $M\models h(\vec{a},0)=b$.

(b) Fix $\vec{a},b,c,d\in M$. To prove $PA\vdash\forall \vec{x},y,z,z'(h(\vec{x},y)=z\wedge g(\vec{x},y,z)=z' \leftrightarrow h(\vec{x},y+1) = z' )$, by completeness of first order logic, it suffices to prove that $M\models h(\vec{a},b)=c \wedge g(\vec{a},b,c)=d \leftrightarrow h(\vec{a},b+1)=d$. Consider $M\models h(\vec{a},b)=c \wedge g(\vec{a},b,c)=d$ as follows: by definition of $h$, it is equivalent to saying that there exists $w\in M$ such that $M \models f(\vec{a})=(w)_0 \wedge \forall i<b\ g(\vec{a},i,(w)_i) = (w)_{i+1} \wedge (w)_b = c$ and $M \models g(\vec{a},b,c) = d$; by Lemma \ref{beta_function} (b), it is equivalent to saying that there exist $w,w'\in M$ such that $M \models f(\vec{a})=(w)_0 \wedge \forall i<b\ g(\vec{a},i,(w)_i)=(w)_{i+1} \wedge (w)_b = c$, $M \models g(\vec{a},b,c) = d$ and $M \models \forall i < b+1\ (w')_i=(w)_i \wedge (w')_{b+1} = d$; letting $w=w'$, it is equivalent to saying that there exists $w'\in M$ such that $M \models f(\vec{a})=(w')_0 \wedge \forall i < b+1\ g(\vec{a},i,(w')_i) = (w')_{i+1} \wedge (w')_{b+1} = d$; by definition of $h$, it is equivalent to $M\models h(\vec{a},b+1) = d$.

(c) Fix $\vec{a},b,c,d\in M$. To prove $PA\vdash\forall \vec{x},y,z,z'(h(\vec{x},y)=z\wedge h(\vec{x},y)=z'\rightarrow z=z')$, by completeness of first order logic, it suffices to prove that $M\models h(\vec{a},b)=c\wedge h(\vec{a},b)=d \rightarrow c=d$. Suppose that $M\models h(\vec{a},b)=c$ and $M\models h(\vec{a},b)=d$. Then we have to prove that $c=d$. By the supposition, there exist $w,w'\in M$ such that $M\models f(\vec{a})=(w)_0\wedge \forall i<b\ g(\vec{a},i,(w)_i)=(w)_{i+1} \wedge (w)_b=c$ and $M \models f(\vec{a})=(w')_0\wedge \forall i<b\ g(\vec{a},i,(w')_i) =(w')_{i+1} \wedge (w')_b=d$. It's trivial that $M\models (w)_0 = (w')_0$. For any $i<b$, $M\models (w)_i=(w')_i$ implies $M\models (w)_{i+1}=(w')_{i+1}$, since $g$ is a function. By induction on $i$ up to $b$, it follows that $M\models \forall i\leq b\ (w)_i=(w')_i$. In particular it follows that $(w)_b = (w')_b$ and finally we have that $c=d$.
\end{proof}

For a more suggestive purpose, we often write the function $h$ defined by recursion from functions $f$ and $g$ as the following:
\begin{equation*}
\left\{
  \begin{array}{l}
    h(\vec{x},0) = f(\vec{x}); \\
    h(\vec{x},y+1) = g(\vec{x},y,h(\vec{x},y)).
  \end{array}
\right.
\end{equation*}

\theoremstyle{definition}
\newtheorem{minimization}[composition]{Definition}
\begin{minimization}[Minimization]
Let $\varphi(\vec{x},y,z)$ be an $L$-formula such that $PA\vdash \forall \vec{x},y,z,z'(\varphi(\vec{x},y,z)\wedge \varphi(\vec{x},y,z')\rightarrow z=z')$, $\varphi$ defining the function $z = f(\vec{x},y)$. Define $y=h(\vec{x})$ from $f$ by the $L$-formula $\theta(\vec{x},y)$, i.e. $\varphi(\vec{x},y,0) \wedge \forall i<y \exists z ( \varphi(\vec{x},i,z) \wedge z \neq 0)$. The process defined by $\theta$ from $f$ is called minimization.
\end{minimization}
\theoremstyle{plain}
\newtheorem{well_defined_minimization}[composition]{Lemma}
\begin{well_defined_minimization}\label{well_defined_minimization}
Let $h$ be as defined before. Then we have that $PA \vdash \forall \vec{x},y,y'(h(\vec{x})=y\wedge h(\vec{x})=y'\rightarrow y=y')$.
\end{well_defined_minimization}
\begin{proof}
Let $M\models PA$. Fix $\vec{a},b,c\in M$. To prove $PA \vdash \forall \vec{x},y,y'(h(\vec{x})=y\wedge h(\vec{x})=y'\rightarrow y=y')$, by completeness of first order logic, it suffices to prove that $M \models h(\vec{a})=b\wedge h(\vec{a})=c\rightarrow b=c$. Suppose that $M\models h(\vec{a})=b$ and $M\models h(\vec{a})=c$. Then we have to prove that $b=c$. Assume for a contradiction that $b \neq c$. By the order relation of $M$ \cite[Section 25.1]{c. and l.}, it follows that $b<c$ or $c<b$. Without loss of generality, suppose that $b<c$. Since $M \models h(\vec{a})=c$, it follows that there exists $d\in M$ such that $M\models f(\vec{a},b)=d \neq 0$, a contradiction to $M\models f(\vec{a},b)=0$, which is an implication of $M\models h(\vec{a})=b$.
\end{proof}

We also put the function $h$ defined by minimization from $f$ as $h(\vec{x}) = \mu y (f(\vec{x},y)=0)$.

\theoremstyle{definition}
\newtheorem{the_definition_of_x_recursive_functions}[composition]{Definition}
\begin{the_definition_of_x_recursive_functions}
A function $h$ of $n$ arguments is X-recursive iff it belongs to one of the following categories.

(1) (Elementals) That is
\begin{equation*}
  h = \lambda \vec{x}.m \mid \lambda \vec{x}.id_i^n(\vec{x})  \mid \lambda x_1,x_2.(x_1+x_2) \mid \lambda x_1,x_2.(x_1\cdot x_2),
\end{equation*}
where $m\in N$ and $id_i^n(x_1,\ldots,x_i,\ldots,x_n)=x_i$.

(2) (Composition) There are X-recursive functions $f$ of $m$ arguments and $g_1,\ldots,g_m$ each of $n$ arguments such that
\begin{equation*}
h(\vec{x}) = f( g_1(\vec{x}), \ldots, g_m(\vec{x}) ).
\end{equation*}
In this case, denote $h$ by $Cn[f,g_1,\ldots,g_m]$.

(3) (Recursion) There are X-recursive functions $f$ of $n-1$ arguments and $g$ of $n+1$ arguments such that
\begin{equation*}
\left\{
  \begin{array}{l}
    h(\vec{x},0) = f(\vec{x}) ; \\
    h(\vec{x},y+1)=g(\vec{x},y,h(\vec{x},y)) .
  \end{array}
\right.
\end{equation*}
In this case, denote $h$ by $Pr[f,g]$.

(4) (Minimization) There is an X-recursive function $f$ of $n+1$ arguments such that
\begin{equation*}
h(\vec{x}) = \mu y (f(\vec{x},y)=0).
\end{equation*}
In this case, denote $h$ by $Mn[f]$.

Functions obtained from the elementary functions only by composition and recursion are called primitive X-recursive.
\end{the_definition_of_x_recursive_functions}

\newtheorem{the_formula_defining_x_recursive_function}[composition]{Definition}
\begin{the_formula_defining_x_recursive_function}
  For every X-recursive function $h$, the generalized $\Sigma_1$-formula $\gamma_h \in L$ is defined inductively as follows.

(1) $\gamma_h(\vec{x},y) ::= y=m \mid y=x_i \mid y=x_1+x_2 \mid y=x_1\cdot x_2$;

(2) $\gamma_h(\vec{x},y) ::= \exists \vec{z}(\bigwedge_{1 \leq i \leq m} \gamma_{g_i}(\vec{x},z_i) \wedge \gamma_f(\vec{z},y))$;

(3) $\gamma_h(\vec{x},y,z) ::= \exists w (\gamma_f(\vec{x},(w)_0) \wedge \forall i<y\ \gamma_g(\vec{x},i,(w)_i, (w)_{i+1}) \wedge (w)_y = z)$;

(4) $\gamma_h(\vec{x},y) ::= \gamma_f(\vec{x},y,0) \wedge \forall i<y \exists z(\gamma_f(\vec{x},i,z)\wedge z\neq 0)$.
\end{the_formula_defining_x_recursive_function}

Summarizing the above, we formulate that
\theoremstyle{plain}
\newtheorem{representability_of_X_recursive_functions}[composition]{Theorem}
\begin{representability_of_X_recursive_functions}[Representability of X-recursive functions]\label{representability_of_X_recursive_functions}
For every X-recursive function $h$, $PA$ proves that

(a) $\forall \vec{x},y ( h(\vec{x})=y \leftrightarrow \gamma_h(\vec{x},y) )$;

(b) $\forall \vec{x},y,y' (\gamma_h(\vec{x},y)\wedge \gamma_h(\vec{x},y')\rightarrow y=y')$.
\end{representability_of_X_recursive_functions}

Note that $\{h^N : h \mbox{ is an X-recursive function }\}$, where $h^N$ is the denotation of $h$ in $N$, is precisely the set of recursive functions \cite{rogers_1}. Thus, X-recursive functions are generalizations of the classical recursive functions from the standard structure to nonstandard models of $PA$ with the uniform $\Sigma_1$-definability.

\subsection{Properties of X-recursive functions}\label{properties_of_X_recursive_functions}

\theoremstyle{definition}
\newtheorem{X_recursive_relations}{Definition}[subsection]
\begin{X_recursive_relations}\label{X_recursive_relations}
For any $n$-place function $f$ and relation $R$ defined in $PA$, $f$ is the characteristic function of $R$ iff $PA \vdash \forall \vec{x}(R(\vec{x})\rightarrow f(\vec{x})=1 \wedge \neg R(\vec{x})\rightarrow f(\vec{x})=0)$; the characteristic function of $R$ is usually denoted $\chi_R$. A relation $R$ is (primitive) X-recursive iff $\chi_R$ is (primitive) X-recursive.
\end{X_recursive_relations}

\theoremstyle{plain}
\newtheorem{sigma_0_is_primitive_X_recursive}[X_recursive_relations]{Lemma}
\begin{sigma_0_is_primitive_X_recursive}\label{sigma_0_is_primitive_X_recursive}
Every $\Sigma_0$-formula (or $\Pi_0$-formula) defines a primitive X-recursive relation.
\end{sigma_0_is_primitive_X_recursive}
\begin{proof}
  The proof technique for every $\Sigma_0$-formula defining a primitive recursive relation has been shown in \cite[Section 7.1]{c. and l.}, and it also hold for nonstandard models. For more details, the reader refers to Appendix \ref{proof_of_sigma_0_being_primitive_X_recursive}.
\end{proof}

\newtheorem{minimization_to_X_recusive_relation}[X_recursive_relations]{Lemma}
\begin{minimization_to_X_recusive_relation}\label{minimization_to_X_recusive_relation}
Let $R$ be an $(n+1)$-place X-recursive relation. Define a total or partial function $r$ by $r(\vec{x}) = \mbox{\ the least $y$ such that\ } R(\vec{x},y)$. Then $r$ is X-recursive.
\end{minimization_to_X_recusive_relation}
\begin{proof}
The proof technique for the classical counterpart in recursion theory is shown in \cite[Proposition 7.9]{c. and l.}, and it also holds for nonstandard models. For more details, the reader refers to Appendix \ref{proof_of_minimization_to_X_recusive_relation}.
\end{proof}

\theoremstyle{plain}
\newtheorem{sigma_1_defines_X_recusive_function}[X_recursive_relations]{Theorem}
\begin{sigma_1_defines_X_recusive_function}\label{sigma_1_defines_X_recusive_function}
Every $\Sigma_1$-formula $\varphi(\vec{x},y)\in L$ with $PA\vdash \forall \vec{x},y,z (\varphi(\vec{x},y) \wedge \varphi(\vec{x},z)\rightarrow y = z)$ defines an X-recursive function $y = f_\varphi(\vec{x})$.
\end{sigma_1_defines_X_recusive_function}
\begin{proof}
 Let $\varphi(\vec{x},y)$ be a $\Sigma_1$-formula with $PA\vdash \forall \vec{x},y,z (\varphi(\vec{x},y) \wedge \varphi(\vec{x},z)\rightarrow y = z)$ and $PA\vdash \forall \vec{x},y ( \varphi(\vec{x},y)\leftrightarrow f_\varphi(\vec{x}) = y )$. Then we have to prove that $y = f_\varphi(\vec{x})$ is X-recursive. By definition of $\Sigma_1$, there exists a $\Sigma_0$-formula $\psi(\vec{x},y,z)\in L$ such that $PA\vdash \forall \vec{x},y (\varphi(\vec{x},y)\leftrightarrow \exists z\ \psi(\vec{x},y,z) )$.

  We now introduce two auxiliary functions defined in $PA$:
  \begin{equation*}
    g(\vec{x}) = \mbox{\ the least $w$ such that\ } \exists y<w \exists z<w\ \psi(\vec{x},y,z),
  \end{equation*}
  \begin{equation*}
    h(\vec{x},y) = \mbox{\ the least $w$ such that\ } w<y \wedge \exists z<y\ \psi(\vec{x},w,z).
  \end{equation*}

  By definition of $\Sigma_0$, one can see that $\exists y<w \exists z<w\ \psi(\vec{x},y,z)\in \Sigma_0$ and $w<y \wedge \exists z<y\ \psi(\vec{x},w,z)\in \Sigma_0$. By Lemma \ref{sigma_0_is_primitive_X_recursive}, it follows that each of $\exists y<w \exists z<w\ \psi(\vec{x},y,z)$ and $w<y \wedge \exists z<y\ \psi(\vec{x},w,z)$ defines a (primitive) X-recursive relation. By Lemma \ref{minimization_to_X_recusive_relation}, it follows that $g$ and $h$ are X-recursive. It's easy to check that $f_\varphi(\vec{x}) = h(\vec{x},g(\vec{x})) = h(id^n_1(\vec{x}),\ldots,id^n_n(\vec{x}),g(\vec{x}))$. Thus $f_\varphi = Cn[h, id^n_1, \ldots, id^n_n, g]$ is X-recursive.
\end{proof}

\subsection{Relationship of $PA^*$, $PA^+$ and $PA\cup Tr^N(\Pi_1)$}\label{relationship_of_PA_star_with_PA_plus}

In the subsection, for while-programs, we should distinguish between the input variables and non-input variables. Let $S\in WP$ have the program variables $\vec{x}=(\vec{p},\vec{q})$ with $\vec{p}$ and $\vec{q}$ being the vectors of input and non-input variables respectively. Define $\alpha_S^{(i)}(\vec{p},y)$ by
\begin{equation*}
\alpha_S^{(i)}(\vec{p},y) ::= \exists \vec{q}, \vec{y} (\alpha_S(\vec{x}, \vec{y}) \wedge y=y_i ).
\end{equation*}
Note that in $\alpha_S^{(i)}(\vec{p},y)$, $y$ is the designated output variable.

\newtheorem{from_X_recursive_functions_to_while_programs}{Lemma}[subsection]
\begin{from_X_recursive_functions_to_while_programs}\label{from_X_recursive_functions_to_while_programs}
For every X-recursive function $h$, there exists $S\in WP$ such that $PA \vdash \forall \vec{p},y(\alpha_S^{(1)}(\vec{p},y) \leftrightarrow \gamma_h(\vec{p},y))$.
\end{from_X_recursive_functions_to_while_programs}
\begin{proof}
  It follows from recursion theory that for every X-recursive function $h$, there exists $S\in WP$ such that $N \models \forall \vec{p},y(\alpha_S^{(1)}(\vec{p},y) \leftrightarrow \gamma_h(\vec{p},y) )$; for nonstandard models $M$ of $PA$, it also holds that $M \models \forall \vec{p},y(\alpha_S^{(1)}(\vec{p},y) \leftrightarrow \gamma_h(\vec{p},y)$; this lemma follows from completeness of first-order logic.
\end{proof}

\newtheorem{sigma_1_defines_a_while_program}[from_X_recursive_functions_to_while_programs]{Lemma}
\begin{sigma_1_defines_a_while_program}\label{sigma_1_defines_a_while_program}
For every $\Sigma_1$-formula $\varphi(\vec{x},y)$ with $PA\vdash \forall \vec{x},y,z (\varphi(\vec{x},y) \wedge \varphi(\vec{x},z)\rightarrow y = z)$, there exists $S\in WP$ such that $PA \vdash \forall \vec{p},y(\alpha_S^{(1)}(\vec{p},y) \leftrightarrow \varphi(\vec{p},y))$.
\end{sigma_1_defines_a_while_program}
\begin{proof}
  Immediate from Theorem \ref{sigma_1_defines_X_recusive_function} and Lemma \ref{from_X_recursive_functions_to_while_programs}.
\end{proof}

\theoremstyle{plain}
\newtheorem{equal_sets_of_theorems}[from_X_recursive_functions_to_while_programs]{Theorem}
\begin{equal_sets_of_theorems}\label{equal_sets_of_theorems}
  $Thm(PA^*) = Thm(PA^+) = Thm(PA\cup Tr^N(\Pi_1))$.
\end{equal_sets_of_theorems}
\begin{proof}
By Lemmas \ref{property_of_pa_star} and \ref{property_of_pa_plus}, it follows that $Thm(PA^*)$, $Thm(PA^+)$ $\subseteq$ $Thm(PA\cup Tr^N(\Pi_1))$. Then we have to prove that $Thm(PA^*)$, $Thm(PA^+)$ $\supseteq$ $Thm(PA\cup Tr^N(\Pi_1))$. It suffices to prove that $PA^*\vdash Tr^N(\Pi_1)$ and $PA^+\vdash Tr^N(\Pi_1)$. Fix $\varphi\in Tr^N(\Pi_1)$. It remains to show that $PA^*\vdash \varphi$ and $PA^+\vdash \varphi$. By definition of $Tr^N(\Pi_1)$, there exists a $\Sigma_0$-formula $\psi(y)$ such that $\varphi \equiv \forall y\ \psi(y)$ and $N\models \forall y\ \psi(y)$. Define $\phi(x,y)\in\Sigma_0$ by $\phi(x,y) ::= x = x \wedge \neg\psi(y)\wedge \forall i<y\ \psi(i)$. By the least number principle, it follows that $PA\vdash \exists y\ \neg \psi(y) \leftrightarrow \exists y(\neg\psi(y)\wedge \forall i<y\ \psi(i))$. Negating both sides of $\leftrightarrow$, we have that $PA\vdash \forall y\ \psi(y) \leftrightarrow \forall y \neg (\neg\psi(y)\wedge \forall i<y\ \psi(i))$. By inserting the valid formula $x = x$ into the right side of $\leftrightarrow$, it follows that $PA\vdash \forall y\ \psi(y) \leftrightarrow \forall y \neg (x = x \wedge \neg\psi(y)\wedge \forall i<y\ \psi(i))$. By definition of $\varphi$ and $\phi$, it follows that $PA\vdash \varphi\leftrightarrow \forall y\neg\phi(x,y)$. On the other hand, it's easy to see that $PA\vdash \forall x,y,z (\phi(x,y) \wedge \phi(x,z) \rightarrow y = z )$. By Lemma \ref{sigma_1_defines_a_while_program},  there exists $S\in WP$ such that $PA \vdash \forall x,y(\alpha_S^{(1)}(x,y) \leftrightarrow \phi(x,y))$. Then $PA \vdash \forall y\neg\alpha_S^{(1)}(x,y) \leftrightarrow \forall y\neg\phi(x,y)$ follows. Since $PA\vdash \varphi\leftrightarrow \forall y\neg\phi(x,y)$, we have that $PA \vdash \varphi \leftrightarrow \forall y\neg\alpha_S^{(1)}(x,y)$. By definition of $\alpha_S^{(1)}(x,y)$ (note that $\vec{p} = x$), it follows that $PA \vdash \varphi \leftrightarrow \forall \vec{x}, \vec{y}\neg\alpha_S(\vec{x},\vec{y})$. By soundness of first-order logic, it follows that $N \models \varphi \leftrightarrow \forall \vec{x}, \vec{y}\neg\alpha_S(\vec{x},\vec{y})$. Since $N\models \varphi$, we have that $N\models \forall \vec{x}, \vec{y}\neg\alpha_S(\vec{x},\vec{y})$. By definition of $PA^*$, it follows that $PA^*\vdash \forall \vec{x}, \vec{y}\neg\alpha_S(\vec{x},\vec{y})$. Since $PA \vdash \varphi \leftrightarrow \forall \vec{x}, \vec{y}\neg\alpha_S(\vec{x},\vec{y})$, we have that $PA^* \vdash \varphi \leftrightarrow \forall \vec{x}, \vec{y}\neg\alpha_S(\vec{x},\vec{y})$. Then $PA^*\vdash \varphi$ follows. Fix $\vec{n}\in N$. Since $N\models \forall \vec{x}, \vec{y}\neg\alpha_S(\vec{x},\vec{y})$, we have that $N\models \forall \vec{y}\neg\alpha_S(\vec{n},\vec{y})$. By definition of $PA^+$, it follows that $PA^+\vdash \forall \vec{y}\neg\alpha_S(\vec{n},\vec{y})$. Since $PA \vdash \varphi \leftrightarrow \forall \vec{y}\neg\alpha_S(\vec{n},\vec{y})$, we have that $PA^+ \vdash \varphi \leftrightarrow \forall \vec{y}\neg\alpha_S(\vec{n},\vec{y})$. Then $PA^+\vdash \varphi$ follows.
\end{proof}

\section{Discussion of the results}

In this paper, by including nonstandard inputs, we have shown that $PA^*$, or equivalently $PA\cup Tr^N(\Pi_1)$, is the minimal extension $T$ of $PA$ such that $HL(T)$ is complete relative to $N$ for $\{Cnt\}WP\{Cnt\}$ with nonstandard inputs. We have shown that for any $i>0$, $HL(PA\cup Tr^N(\Pi_i))$ is complete relative to $N$ for $\{P\}WP\{Q\}$ with nonstandard inputs iff $P\subseteq \Sigma_i$ and $Q\subseteq \Pi_i$; and if $\Pi_{i-1} \subseteq P \subseteq \Sigma_i$ or $\Sigma_{i-1} \subseteq Q \subseteq \Pi_i$, then $HL(T)$ is complete relative to $N$ for $\{P\}WP\{Q\}$ with nonstandard inputs iff $Thm(T) \supseteq Thm(PA \cup Tr^N(\Pi_i))$.

By excluding nonstandard inputs, we have shown that $PA^+$, or equivalently $PA\cup Tr^N(\Pi_1)$, is the minimal extension $T$ of $PA$ such that $HL(T)$ is complete relative to $N$ for $\{Cnt\}WP\{Cnt\}$ without nonstandard inputs. We have shown that for any $i>0$, $HL(PA\cup Tr^N(\Pi_i))$ is complete relative to $N$ for $\{P\}WP\{Q\}$ without nonstandard inputs iff $P\subseteq \Pi_{i+1}$ and $Q\subseteq \Sigma_{i+1}$; and if $\Sigma_i \subseteq P \subseteq \Pi_{i+1}$ or $\Pi_i \subseteq Q \subseteq \Sigma_{i+1}$, then $HL(T)$ is complete relative to $N$ for $\{P\}WP\{Q\}$ without nonstandard inputs iff $Thm(T) \supseteq Thm(PA \cup Tr^N(\Pi_i))$.

Observe from the above results that in $HL(PA\cup Tr^N(\Pi_i))$, by excluding nonstandard inputs, the admissible maximal scope of preconditions and postconditions is extended from pre-$\Sigma_i$ and post-$\Pi_i$ to pre-$\Pi_{i+1}$ and post-$\Sigma_{i+1}$; the minimal scope of preconditions and postconditions upon which the full theory of $PA\cup Tr^N(\Pi_i)$ acts is extended from pre-$\Pi_{i-1}$ and post-$\Sigma_{i-1}$ to pre-$\Sigma_i$ and post-$\Pi_i$; yet this restriction has no effects on the completeness of $HL(T)$ relative to $N$ for $\{Cnt\}WP\{Cnt\}$: $PA \cup Tr^N(\Pi_1)$ is minimal in both cases. Considering $Thm(PA)\subsetneqq Thm(PA\cup Tr^N(\Pi_i)) \subsetneqq Th(N)$ and $Th(N) = \bigcup_{i=1}^{\infty}Thm(PA\cup Tr^N(\Pi_i))$, the completeness gap between $HL(PA)$ and $HL(Th(N))$ has been bridged.

Cook's completeness result allows for the whole set of arithmetical formulas as assertions, at the price of using $Th(N)$ as an oracle for the assertion theory. By restricting assertions to subclasses of arithmetical formulas, we show that arithmetical extensions of $PA$ suffice to act as the assertion theory, and the lower the  level of the assertions in the arithmetical hierarchy the lower the level of the required assertion theory is. In conclusion, our completeness results refine Cook's one by reducing the complexity of the assertion theory.

\section*{Acknowledgement}
The authors would like to thank the 973 Program of China (Grant No. 2014CB340701), the National Natural Science Foundation of China (Grant Nos. 61672504 and 61472474), and the CAS-SAFEA International Partnership Program for Creative Research Teams for the financial support.

\appendix

\section{On X-recursion theory}

\subsection{Examples of X-recursive functions}

\theoremstyle{definition}
\newtheorem{summation_product}{Example}[subsection]
\begin{summation_product}[The summation and product functions] \label{summation_product_label}
Let $f$ be a (primitive) X-recursive function of $n+1$ arguments. Then the following functions
\begin{equation*}
   \left\{
     \begin{array}{l}
       g(\vec{x},0)=f(\vec{x},0) \\
       g(\vec{x},y+1)=f(\vec{x},y+1)+g(\vec{x},y)
     \end{array}
   \right.
\end{equation*}
and
\begin{equation*}
   \left\{
     \begin{array}{l}
       h(\vec{x},0)=f(\vec{x},0) \\
       h(\vec{x},y+1)=f(\vec{x},y+1)\cdot h(\vec{x},y)
     \end{array}
   \right.
\end{equation*}
are (primitive) X-recursive. Intuitively, we put $g$ and $h$ in the following form: $g(\vec{x},y) = f(\vec{x},0) + f(\vec{x},1) + \ldots + f(\vec{x},y) = \sum_{i=0}^{y} f(\vec{x},i)$ and $h(\vec{x},y) = f(\vec{x},0) \cdot f(\vec{x},1) \cdot \ldots \cdot f(\vec{x},y) = \prod_{i=0}^{y} f(\vec{x},i)$.
\end{summation_product}
\begin{proof}
Strictly speaking, $g$ has the following form
\begin{multline*}
  Pr[ Cn[f,id_1^n,\ldots,id_n^n,0], Cn[+,Cn[f,id_1^{n+2}, \\
  \ldots,id_n^{n+2},Cn[+,id_{n+1}^{n+2},1]],id_{n+2}^{n+2}] ].
\end{multline*}
Similarly for $h$.
\end{proof}

In the above, $\gamma_g$ and $\gamma_h$ have the following logically equivalent forms respectively: $\exists w ( f(\vec{x},0)=(w)_0 \wedge \forall i<y\ f(\vec{x},i+1)+(w)_i = (w)_{i+1} \wedge (w)_y = z )$ and $\exists w ( f(\vec{x},0)=(w)_0 \wedge \forall i<y\ f(\vec{x},i+1)\cdot(w)_i = (w)_{i+1} \wedge (w)_y = z )$. In the following, for simplicity, we often write the construction processes of X-recursive functions in an informal style.

\newtheorem{simple_X_recursive_functions}[summation_product]{Example}
\begin{simple_X_recursive_functions}\label{simple_X_recursive_functions_label}
The following functions are (primitive) X-recursive:

(a) (The predecessor function) Define $pred(x)$ to be the predecessor $x-1$ of $x$ for $x>0$, and let $pred(0)=0$ by convention.

(b) (The difference function) Define $x-y$ to be $z$ such that $x = y+z$ if $x\geq y$, and let $x-y = 0$ by convention otherwise.

(c) (The signum functions) Define $sg(0)=0$ and $sg(x)=1$ if $x>0$, and define $\overline{sg}(0)=1$ and $\overline{sg}(x)=0$ if $x>0$.
\end{simple_X_recursive_functions}
\begin{proof}
(a) Define the primitive X-recursive function $p$ as follows:
\begin{equation*}
\left\{
  \begin{array}{l}
    p(0)=0; \\
    p(x+1)=x.
  \end{array}
\right.
\end{equation*}
To show $pred = p$, by Theorem \ref{representability_of_X_recursive_functions}, it suffices to prove that $PA \vdash \forall x,y ( pred(x)=y \leftrightarrow \gamma_p(x,y) )$. Fix $M\models PA$. By completeness of first order logic, it suffices to prove that $M \models \forall x,y ( pred(x)=y \leftrightarrow \gamma_p(x,y) )$ by induction on $x$. For $x=0$, consider $M\models pred(0)=y$ as follows: by definition of $pred$, it is equivalent to $M\models y=0$; by Lemma \ref{beta_function} (a), it is equivalent to $M\models \exists w ( (w)_0=0 \wedge (w)_0=y )$; by pure logic, it is equivalent to $M\models \exists w ( (w)_0=0 \wedge \forall i < 0\ (w)_{i+1} = i \wedge (w)_0=y )$; by definition of $\gamma_p$, it is equivalent to $M \models \gamma_p(0,y)$. As the inductive hypothesis, suppose that $M \models \forall y ( pred(x)=y \leftrightarrow \gamma_p(x,y) )$ for any $x=a\in M$. Then we have to prove that $M \models \forall y ( pred(x)=y \leftrightarrow \gamma_p(x,y) )$ for $x=a+1$. Consider $M \models  pred(a+1)=y$ as follows: by definition of $pred$, it is equivalent to $M \models y=a$; by the induction hypothesis, it is equivalent to saying that there exists $w\in M$ such that $M \models (w)_0=0 \wedge \forall i < a\ (w)_{i+1} = i \wedge (w)_a = pred(a)$, and $M\models y=a$; by Lemma \ref{beta_function} (b), it is equivalent to saying that there exist $w,w'\in M$ such that $M \models (w)_0=0 \wedge \forall i < a\ (w)_{i+1} = i \wedge (w)_a = pred(a)$, $M \models \forall i < a+1\ (w')_i = (w)_i \wedge (w')_{a+1} = a$, and $M \models y=a$; letting $w=w'$, it is equivalent to saying that there exists $w' \in M$ such that $M \models (w')_0=0 \wedge \forall i < a+1\ (w')_{i+1} = i \wedge (w')_{a+1} = y$; by pure logic, it is equivalent to $M \models \exists w( (w)_0=0 \wedge \forall i < a+1\ (w)_{i+1} = i \wedge (w)_{a+1} = y )$; by definition of $\gamma_p$, it is equivalent to $M \models \gamma_p(a+1,y)$.

(b) Define the primitive X-recursive function $x \ominus y$ as follows:
\begin{equation*}
\left\{
  \begin{array}{l}
    x \ominus 0=x; \\
    x \ominus (y+1)=pred(x \ominus y).
  \end{array}
\right.
\end{equation*}
To show $- = \ominus$, by Theorem \ref{representability_of_X_recursive_functions}, it suffices to prove that $PA \vdash \forall x,y,z ( x - y=z \leftrightarrow \gamma_{\ominus}(x,y,z) )$. Fix $M\models PA$. By completeness of first order logic, it suffices to prove that $M \models \forall x,y,z ( x - y=z \leftrightarrow \gamma_{\ominus}(x,y,z) )$. Fix $a\in M$. It suffices to prove that $M \models a - y=z$ iff $M \models \gamma_{\ominus}(a,y,z)$ by induction on $y$. For $y=0$, consider $M \models a - 0 = z$ as follows: by definition of $-$, it is equivalent to $M \models z=a$; by Lemma \ref{beta_function} (a), it is equivalent to $M\models \exists w ( (w)_0=a \wedge (w)_0=z )$; by pure logic, it is equivalent to $M \models \exists w ( (w)_0=a \wedge \forall i < 0\ pred((w)_{i}) = (w)_{i+1} \wedge (w)_0 = z )$; by definition of $\gamma_{\ominus}$, it is equivalent to $M \models \gamma_{\ominus}(a,0,z)$. As the inductive hypothesis, suppose that $M \models a - y=z \leftrightarrow \gamma_{\ominus}(a,y,z)$ for any $y=b\in M$. Then we have to prove that $M \models a - y=z \leftrightarrow \gamma_{\ominus}(a,y,z)$ for $y = b+1$. Consider $M \models a - (b+1)=z$ as follows: by pure logic, it is equivalent to saying that for some $u\in M$, $M \models a - b=u$ and $M\models pred(u)=z$; by the induction hypothesis, it is equivalent to saying that for some $u\in M$, $M \models \gamma_{\ominus}(a,b,u)$ and $M \models pred(u)=z$; by definition of $\gamma_{\ominus}$, it is equivalent to saying that there exist $u,w\in M$ such that $M \models (w)_0 = a \wedge \forall i < b\ pred((w)_{i}) = (w)_{i+1} \wedge (w)_b = u$ and $M\models pred(u)=z$; by Lemma \ref{beta_function} (b), it is equivalent to saying that there exist $u,w,w'\in M$ such that $M \models (w)_0 = a \wedge \forall i < b\ pred((w)_{i}) = (w)_{i+1} \wedge (w)_b = u$, $M \models \forall i < b+1\ (w')_i = (w)_i \wedge (w')_{b+1} = pred(u)$ and $M \models pred(u)=z$; letting $w=w'$, it is equivalent to saying that there exists $w'\in M$ such that $M \models (w')_0 = a \wedge \forall i < b+1\ pred((w')_{i}) = (w')_{i+1} \wedge (w')_{b+1} = z$; by pure logic, it is equivalent to $M \models \exists w( (w)_0 = a \wedge \forall i < b+1\ pred((w)_{i}) = (w)_{i+1} \wedge (w)_{b+1} = z )$; by definition of $\gamma_{\ominus}$, it is equivalent to $M \models \gamma_{\ominus}(a,b+1,z)$.

(c) It's easy to check that $sg(x) = 1 - (1 - x)$ and $\overline{sg}(x) = 1- x$.
\end{proof}

\subsection{Examples and properties of X-recursive relations}

\newtheorem{identity_and_order}{Example}[subsection]
\begin{identity_and_order}[Identity and order]\label{identity_order_label}
The identity relation, which holds if and only if $x=y$, is primitive X-recursive, since a little thought shows its characteristic function $\chi_{=}(x,y)$ is defined by $\chi_{=}(x,y) ::= 1 - ( sg(x - y)+sg(y - x))$. The strict less-than order relation, which holds if and only if $x<y$, is also primitive X-recursive, since its characteristic function $\chi_{<}(x,y)$ is defined by $\chi_{<}(x,y) ::= sg(y - x)$.
\end{identity_and_order}

We are now ready to indicate an important process for obtaining new (primitive) X-recursive functions from old.

\theoremstyle{plain}
\newtheorem{lemma_relation_2}[identity_and_order]{Lemma}
\begin{lemma_relation_2}[Definition by cases] \label{definition_by_cases_label}
Suppose that $f$ is the function defined in the following form:
\begin{equation*}
  f(\vec{x}) = \left\{
             \begin{array}{ll}
               g_1(\vec{x}) & \hbox{if $C_1(\vec{x})$} \\
               \vdots & \hbox{$\vdots$} \\
               g_n(\vec{x}) & \hbox{if $C_n(\vec{x})$}
             \end{array}
           \right.
\end{equation*}
where $C_1,\ldots,C_n$ are (primitive) X-recursive relations that are mutually exclusive and collectively exhaustive, and where $g_1,\ldots,g_n$ are (primitive) X-recursive functions. Then $f$ is (primitive) X-recursive.
\end{lemma_relation_2}
\begin{proof}
Let $c_i$ be the characteristic function of $C_i$. Define $h$ as follows:
\begin{equation*}
  h(\vec{x}) = \sum_{i=1}^{n} g_i(\vec{x})\cdot c_i(\vec{x}).
\end{equation*}
The function $h$ is (primitive) X-recursive since it is obtainable by compositions from the $g_i$ and $c_i$, which are (primitive) X-recursive by assumption, together with the addition and multiplication (and identity) functions. It's easy to verify that $f=h$.
\end{proof}

\theoremstyle{definition}
\newtheorem{example_relation_3}[identity_and_order]{Example}
\begin{example_relation_3}[The maximum and minimum functions]
As an example of definition by cases, consider $max(x,y) = $ the larger of the numbers $x, y$. This can be defined as follows:
\begin{equation*}
max(x,y) = \left\{
             \begin{array}{ll}
               x & \hbox{if $x \geq y$} \\
               y & \hbox{if $x<y$}
             \end{array}
           \right.
\end{equation*}
or in the official format of the lemma above with $g_1=id_1^2$ and $g_2=id_2^2$. Similarly, function $min(x,y) = $ the smaller of $x,y$ is also primitive X-recursive.
\end{example_relation_3}

Besides definition by cases, there are a variety of processes for defining new relations from old that can be shown to produce new (primitive) X-recursive relations when applied to (primitive) X-recursive relations. The following theorem is stated for X-recursive relations (and total X-recursive functions), but hold equally for primitive X-recursive relations (and primitive X-recursive functions), by the same proofs, though it would be tedious for writers and readers alike to include a bracketed `(primitive)' everywhere in the statement and proof of the result.

\theoremstyle{plain}
\newtheorem{lemma_relation_4}[identity_and_order]{Theorem}
\begin{lemma_relation_4}[Closure properties of X-recursive relations]\label{closure_properties_label}

(a) A relation defined by substituting total X-recursive functions in an X-recursive relation is X-recursive.

(b) The graph relation of any total X-recursive function is X-recursive.

(c) If a relation is X-recursive, so is the relation defined by its negation.

(d) If two relations are X-recursive, then so is the relation defined by their conjunction.

(e) If two relations are X-recursive, then so is the relation defined by their disjunction.

(f) If a relation is X-recursive, then so is the relation defined from it by bounded universal quantification.

(g) If a relation is X-recursive, then so is the relation defined from it by bounded existential quantification.

\end{lemma_relation_4}

\begin{proof}

(a), (b): Trivially.

(c): The characteristic function $c^*$ of the negation or complement of $R$ is obtainable from the characteristic function $c$ of $R$ by $c^*(\vec{x}) ::= 1 - c(\vec{x})$.

(d), (e): The characteristic function $c^*$ of the conjunction or intersection of $R_1$ and $R_2$ is obtainable from the characteristic functions $c_1$ and $c_2$ of $R_1$ and $R_2$ by $c^*(\vec{x}) ::= min(c_1(\vec{x}),c_2(\vec{x}))$, and the characteristic function $c^\dag$ of the disjunction or union is similarly obtainable using $max$ in place of $min$.

(f): From the characteristic function $c(\vec{x},y)$ of the relation $R(\vec{x},y)$, we define the following X-recursive function:
\begin{equation*}
  u(\vec{x},y) = \prod_{i=0}^{y} c(\vec{x},i),
\end{equation*}
where the product ($\prod$) notation is defined as in Example \ref{summation_product_label}. It remains to prove that $u$ is the characteristic function of the relation $\forall v\leq y\ R(\vec{x},v)$. Fix $\vec{a},b\in M \models PA$. It suffices to prove that $M \models u(\vec{a},b) = 1$ iff $M \models \forall i\leq b\ R(\vec{a},i)$. Consider $M \models u(\vec{a},b) = 1$ as follows: by definition, it is equivalent to saying that there exists $w\in M$ such that $M \models c(\vec{a},0)=(w)_0 \wedge \forall i<b\ c(\vec{a},i+1)\cdot (w)_i = (w)_{i+1} \wedge (w)_b = 1$; ($\Rightarrow$. It suffices to prove that $M\models \forall x \leq b\ (w)_{b-x} = 1$ by induction on $x$. For $x=0$, we immediately have $M\models (w)_{b}=1$. As the inductive hypothesis, suppose that $M\models (w)_{b-x}=1$ for any $x=d < b$. Then we have to prove that $M\models (w)_{b-x}=1$ for $x=d+1$. Since $M \models c(\vec{a},b-d)\cdot(w)_{b-(d+1)} = (w)_{b-d}$, it follows that $M \models (w)_{b-(d+1)}=1$.) by the bracketed argument, it is equivalent to $M \models \exists w \forall i \leq b\ (w)_i = c(\vec{a},i) = 1$; ($\Leftarrow$. It suffices to prove that $M \models \forall x \exists w \forall i \leq x\ (w)_i = c(\vec{a},i)$ by induction on $x$. For $x=0$, $M \models \exists w \forall i \leq 0\ (w)_i = c(\vec{a},i)$ follows from Lemma \ref{beta_function} (a). As the inductive hypothesis, suppose that $M \models \exists w \forall i \leq x\ (w)_i = c(\vec{a},i)$ for any $x=b\in M$. Then we have to prove that $M \models \exists w \forall i \leq x\ (w)_i = c(\vec{a},i)$ for $x = b+1$. This is the case due to Lemma \ref{beta_function} (b).) by the bracketed argument, it is equivalent to $M \models \forall i \leq b\ c(\vec{a},i) = 1$; by definition of $c$, it is equivalent to $M \models \forall i\leq b\ R(\vec{a},i)$. For the strict bounds $\forall v<y$ and $\exists v<y$ we need only replace $y$ by $y-1$.

(g) From the characteristic function $c(\vec{x},y)$ of the relation $R(\vec{x},y)$, we define the following X-recursive function:
\begin{equation*}
  e(\vec{x},y) = sg( \sum_{i=0}^{y} c(\vec{x},i) ),
\end{equation*}
where the summation ($\sum$) notation is defined as in Example \ref{summation_product_label}. It remains to prove that $e$ is the characteristic function of the relation $\exists v\leq y\ R(\vec{x},v)$. Fix $\vec{a}, b\in M\models PA$. It suffices to prove that $M \models e(\vec{a},b) = 1$ iff $M \models \exists i \leq b\ R(\vec{a},i)$. Consider $M \models e(\vec{a},b) = 1$ as follows: by definition, it is equivalent to $M \models \exists w ( c(\vec{a},0)=(w)_0 \wedge \forall i<b\ c(\vec{a},i+1)+(w)_i = (w)_{i+1} \wedge sg((w)_b) = 1)$; ($\Leftarrow$. Suppose that $M \not\models sg((w)_b) = 1$. Then $M \models (w)_b = 0$ follows. It's easy to prove that $M \models \forall i \leq b\ (w)_{b-i} = 0$ by induction on $i$. Then it follows that $M \models \forall i \leq b\ c(\vec{a},i) = 0$, a contradiction. $\Rightarrow$. Suppose that $M \not\models \exists i \leq b\ c(\vec{a},i) = 1$. Then $M \models \forall i \leq b\ c(\vec{a},i) = 0$ follows. It's easy to prove that $M \models \forall i \leq b\ (w)_i = 0$ by induction on $i$, a contradiction.) by the bracketed argument, it is equivalent to $M \models \exists w ( c(\vec{a},0)=(w)_0 \wedge \forall i<b\ c(\vec{a},i+1)+(w)_i = (w)_{i+1})$ and $M\models \exists i \leq b\ c(\vec{a},i) = 1$; ($\Leftarrow$. It suffices to prove $M \models \exists w ( c(\vec{a},0)=(w)_0 \wedge \forall i<x\ c(\vec{a},i+1)+(w)_i = (w)_{i+1})$ by induction on $x$. For $x=0$, $M \models \exists w ( c(\vec{a},0)=(w)_0 \wedge \forall i<0\ c(\vec{a},i+1)+(w)_i = (w)_{i+1})$ follows from Lemma \ref{beta_function} (a). As the inductive hypothesis, suppose that $M \models \exists w ( c(\vec{a},0)=(w)_0 \wedge \forall i<x\ c(\vec{a},i+1)+(w)_i = (w)_{i+1})$ for any $x=b\in M$. Then we have to prove that $M \models \exists w ( c(\vec{a},0)=(w)_0 \wedge \forall i<x\ c(\vec{a},i+1)+(w)_i = (w)_{i+1})$ for $x=b+1$. This is the case due to Lemma \ref{beta_function} (b).) by the bracketed argument, it is equivalent to $M \models \exists i \leq b\ c(\vec{a},i) = 1$; by definition of $c$, it is equivalent to $M \models \exists i \leq b\ R(\vec{a},i)$. For the strict bounds $\forall v<y$ and $\exists v<y$ we need only replace $y$ by $y-1$.
\end{proof}

\subsection{Proof of Lemma \ref{sigma_0_is_primitive_X_recursive}} \label{proof_of_sigma_0_being_primitive_X_recursive}

We first show that the boolean relation $B$ is primitive X-recursive by induction on $B$. For $B\equiv E_1 = E_2$ or $B\equiv E_1 < E_2$, $E_1$ and $E_2$ are primitive X-recursive functions by compositions from the elementary functions; it follows from Example \ref{identity_order_label} that $=$ and $<$ are primitive X-recursive relations; since $B$ is the relation defined by substituting primitive X-recursive functions in an primitive X-recursive relation, by Theorem \ref{closure_properties_label} (a), it is primitive X-recursive. For $B\equiv \neg B_1$ or $B\equiv B_1\rightarrow B_2$, by induction hypothesis, $B_1$ and $B_2$ are primitive X-recursive; it follows from Theorem \ref{closure_properties_label} (c) and (d) that $B$ is primitive X-recursive.

We then show that every $\Sigma_0$-formula defines a primitive X-recursive relation. This is the case due to Theorem \ref{closure_properties_label} (f) and (g), together with the fact that the boolean relation $B$ is primitive X-recursive.

\subsection{Proof of Lemma \ref{minimization_to_X_recusive_relation}}\label{proof_of_minimization_to_X_recusive_relation}

Since $R$ is X-recursive, by Theorem \ref{closure_properties_label} (c), it follows that $\neg R$ is X-recursive. The function $r$ is just $Mn[c]$, where $c$ is the characteristic function of $\neg R$.

\end{document}